\documentclass[aps,amsmath,amssymb,twocolumn,showpacs,superscriptaddress]{revtex4}

\usepackage{graphicx}
\usepackage[toc,page]{appendix}
\usepackage{epstopdf}
\usepackage{enumerate}
\usepackage{xcolor}
\usepackage{bm}

\begin{document}

\title{Cotunnelling and polaronic effect in granular systems}
\author{A.S. Ioselevich}
\affiliation{Department of Physics, National Research University ``Higher School of Economics'', Moscow, Russia}
\affiliation{L. D. Landau Institute for Theoretical Physics, Moscow 119334, Russia}
\affiliation{Moscow Institute of Physics and Technology, Moscow 141700, Russia}
\author{V.V. Sivak}
\affiliation{Department of Physics, Yale University, New Haven, Connecticut 06520, USA}
\affiliation{Moscow Institute of Physics and Technology, Moscow 141700, Russia}
\date{\today}

\begin{abstract}
We theoretically study the conductivity in arrays of metallic grains due to the variable-range multiple cotunneling of electrons with short-range (screened) Coulomb interaction. The system is supposed to be coupled to random stray charges in the dielectric matrix that are only loosely bounded to their spatial positions by elastic forces. The flexibility of the stray charges gives rise to a polaronic effect, which leads to the onset of Arrhenius-like conductivity behaviour at low temperatures, replacing conventional Mott variable-range hopping. The effective activation energy logarithmically depends on temperature due to fluctuations of the  polaron barrier heights. We present the unified theory that covers both weak and strong polaron effect regimes of hopping in granular metals and describes the crossover from elastic to inelastic cotunneling.
\end{abstract}


%73.23.-b	Electronic transport in mesoscopic systems
%71.38.-k	Polarons and electron-phonon interactions 
\maketitle

\section{Introduction \label{Introduction}}

In this paper we are discussing a multiparticle cotunneling mechanism of conductivity in a granular metal with flexible charges, randomly placed in the insulating matrix. The flexibility of disorder gives rise to a sort of ``random polaronic effect'', which dramatically affects the temperature dependence of conductivity. A study of this subject requires a combination of different physical concepts and methods, so we start from a brief review of the necessary ingredients. 

\subsection{Variable range hopping  \label{Variable range hopping in disordered }}

The stretched-exponential temperature dependence of conductivity 
\begin{align}
\sigma\propto\exp\{-(T_0/T)^\alpha\}
\label{stre0}
\end{align}
is characteristic of variable range hopping (VRH) in homogeneously disordered materials (such as amorphous solids \cite{mott} or doped semiconductors \cite{ES_book}) at low temperatures. In the case when the long-range interactions do not play any significant role, the exponent $\alpha=1/(d+1)$ (where $d$ is the space dimensionality), $T_0\equiv T_{\rm M}\sim [\nu_F\xi^d]^{-1}$ (where $\nu_F$ is the density of states at the Fermi level and $\xi$ is the inverse decrement of electronic wave-functions), and the corresponding dependence is known as the Mott law \cite{mott0}. In the opposite case, when the long range Coulomb interaction is crucial and gives rise to the soft Coulomb gap in the electronic density of states  \cite{ES_gap},  the exponent $\alpha=1/2$, $T_0\equiv T_{\rm ES}\sim e^2/\kappa \xi$ (where $\kappa$ is the dielectric constant) and this dependence is known as the Efros-Shklovskii law.

\subsection{Variable range cotunneling in granular systems \label{Variable range cotunneling in granular systems}}

A similar behaviour of conductivity  was also observed in  arrays of metallic and semiconducting quantum dots  in the temperature range 1 -- 200K (see reviews \cite{abeles,beloborodov} and some old \cite{VRHold} and more recent \cite{VRH} experimental papers). 
An explanation of  the stretched-exponential $T$-dependence of conductivity in granular materials attracted interest of theorists.
Many of early theories \cite{earlytheory1} were based on special assumptions about the distribution of the random parameters of grains (sizes etc) and were criticised -- see, e.g., \cite{PollakAdkins92}, because of their {\it ad hoc} character and the lack of universality. Some other  theories  \cite{ZhangShklovskii,earlytheory2} correctly indicated the important role of sequential tunneling of electrons through a chain of intermediate grains, but did not give a correct multi-particle description for this tunneling and  a valid recipe for evaluation of the  tunneling amplitude. Such a prescription was worked out in \cite{iosel2005}, where the simple idea, introduced in \cite{ZhangShklovskii}, was generalised to take into account both multi-particle character (i.e., the cotunneling, see \cite{Nazarov1990,NazarovBlanter}) of the process and Coulomb effects. 

In contrast with the standard single-particle-like VRH scenario, where a particle would travel from one end of the chain to the other, consequently hopping through all intermediate grains, the multiple cotunneling scenario, developed in \cite{iosel2005} involves all possible sequences of hops between neighbouring grains in the chain. In general, these hops are executed by different electrons; the intermediate states of the process involve many grains with altered charges. However, upon the completion of the process there is only one net electron that is transferred between the terminal grains of the chain, while  the charges of all intermediate grains return to their initial values. It does not mean that all the processes with different sequences of hops lead to the same final state of the system: the final states of some grains may be identical to the initial ones (elastic cotunneling), while for other grains the initial and final states may differ by an electron-hole pair (inelastic cotunneling). As a result (see \cite{iosel2005} for details), the law \eqref{stre0} is reproduced with
\begin{eqnarray}
T_{0}\sim {\cal L}(T)E_c \label{resu3ggg}
\end{eqnarray}
where
\begin{eqnarray}
 {\cal L}(T)\sim\left\{\begin{aligned}
\ln\left(E_c/g\delta\right),&\quad T\ll T_{c0}\\
\ln\left(E_c^2/gT^2{\cal
L}^2\right),&\quad T\gg T_{c0}
\end{aligned}\right.
\label{resu3}
\end{eqnarray}
and the  Coulomb charging energy is $E_c\sim e^2/2C$, $C$ being typical capacitance in the system of grains. A typical level spacing in a grain $\delta\sim (\nu_Fa^3)^{-1}$, $a$ being the grains size, and $g\ll 1$ is the typical dimensionless conductance between adjacent grains. Note that the logarithmic factor ${\cal L}$ is large.
The crossover between the elastic and inelastic cotunneling regimes takes place at
\begin{eqnarray}
T\sim T_{c0}=\sqrt{E_C\delta}/{\cal L}\label{resu3ppp}
\end{eqnarray}
Thus, the stretched exponential law in the case of granular materials is slightly modified in the intermediate temperature range ($T>T_{c0}$) due to additional logarithmic $T$-dependence of $T_0$. This deviation from the Mott-Efros-Shklovskii law is, however, not easy to detect experimentally.

\subsection{Hard gap and polaronic effect in homogeneously disordered systems \label{Hard gap and polaronic effect in homogeneously disordered systems}}

It is well known that at relatively high temperatures the VRH is not operative, it is changed to the nearest neighbour hopping (NNH), so that the stretched-exponential law \eqref{stre0} is replaced by the Arrhenius law for conductivity $\sigma\propto\exp\{-\varepsilon_3/T\}$ (see \cite{ES_book}). What is much less trivial, in some cases \cite{hard-gap} the reentrance of the Arrhenius law 
\begin{eqnarray}
\sigma\sim\exp\{-E_H/T\}
\label{hard-gap}
\end{eqnarray}
is observed also  at {\it low} temperatures! This reentrance is usually attributed to polaronic effect: a ``hard gap'' $E_H$ is supposed to be related to the energy, necessary for the creation in advance of a polaronic cloud, that then will accommodate a hopping electron at a new position.

 In principle, the hopping electron can take along the necessary energy while hopping from the initial position to the new one, therefore at still lower temperatures the activational mechanism of polaron hopping is substituted by the tunneling one (see \cite{mott}) and the stretched exponential law \eqref{stre0} is again restored at $T\lesssim\omega$, where $\omega$ is the characteristic frequency of phonons (or some other species -- magnons, localised electronic excitations etc), that constitute the polaronic cloud. The case of magnetic polarons is special: due to the local conservation of the magnetisation the process of the tunneling transfer of the polaronic cloud is strongly suppressed  \cite{MHG}, and the classical-quantum crossover is shifted from $T\sim\omega$ to much lower temperatures. It explains why the hard gap phenomenon was experimentally observed predominantly (but not exclusively!) in magnetic systems. In general, the theory of VRH with account for polaronic effect was developed in \cite{MHG,polaron-theory,iossel86,foygel} for different types of polarons.
 
 It is important that, if the strength of the polaronic effect randomly varies from place to place, then it  doesn't necessarily have to lead to the activation Arrhenius law. If the distribution of barrier heights has a power-law tail at zero, one can expect the dependence \eqref{stre0} with $\alpha\neq1$. In particular, for homogeneously disordered solids it was shown in \cite{foygel} that, if the barrier distribution is constant in the vicinity of zero, the Mott conductivity should have the exponent $\alpha=2/(d+2)$.

 \subsection{Polaronic effect in granular systems \label{Polaronic effect in granular systems and its effect on conductivity}}

The onset of Arrhenius-like behaviour of conductivity at low temperatures in granular materials is reported far less often. For example, in  experiments \cite{polaron exp} the Arrhenius behaviour was observed below $\sim5K$ in two-dimensional arrays of semiconducting Ge/Si quantum dots. Above that temperature the conductivity followed Efros-Shklovskii law. In paper \cite{polaron exp 2} the observation of ``almost'' Arrhenius conductivity was reported in the temperature range $20K<T<30K$ in the array of metallic Co nanoparticles. This dependence was well described by \eqref{stre0} with $\alpha=1.1$. At higher temperatures $45K<T<80K$ Efros-Shklovskii law was found.   There are also experiments \cite{polaron exp 3} in which  $\alpha=2/3$ was observed in the broad temperature range $7K<T<200K$ in the arrays of ZnO nanocrystals.  

%With some caution we may hope to explain such results as a manifestation of polaron effect.

It is tempting to attribute these findings to some kind of random polaronic effect. 

But what does the polaron effect mean in the case, where we deal not with single charge carriers, but with grains, containing many electrons? And what are the effective degrees of freedom that constitute here a polaronic cloud? As to our knowledge, no models of the polaronic effect in granular systems have been discussed so far. In this paper we introduce such a model for metallic grains and study its implications for the transport properties within the framework of multiple cotunnelling concept.  

\subsection{Structure of the article \label{Structure of the article}}

The article is organized in the following way. We discuss the basic concepts of our model in Sections \ref{Model} and  \ref{Short range interaction case}. The detailed calculation of electron transition rate between the pair of distant resonant grains in the case of short-range interaction is carried out in Sections \ref{Transition rates}-\ref{physical explanation}. We deal with different parameter ranges and also provide physical interpretation of results. Based on these findings we analyse the temperature behaviour of conductivity in Sections \ref{Conductivity} and \ref{Conductivity  in different temperature ranges}. Section \ref{Summary} contains the summary of our results and in Section \ref{Conclusion} we discuss the limitations and possible future directions of research.

\section{Dynamical fluctuations of the offset charges: General Model \label{Model}}

\begin{figure}[ht]
\includegraphics[width=0.6\columnwidth]{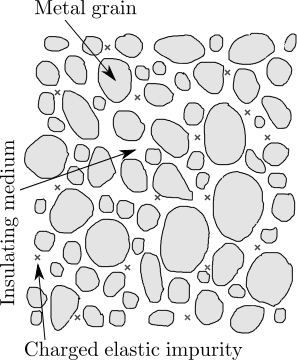}
\caption{Sample of granular material with trapped elastic impurities.} 
\label{grainss}
\end{figure}

The main source of disorder in granular systems is the ``stray charges'' -- hardly removable charged impurities and defects, trapped in the insulating part of the system, see Fig. \ref{grainss}. They produce random Coulomb fields acting on the grains, so that the Coulomb energy of the system is
\begin{align}
E_C(\vec{N},\vec{q})= \frac{1}{2}\left\{(\vec{N}-\vec{Q})\hat{U}(\vec{N}-\vec{Q})-\vec{Q}\hat{U}\vec{Q}\right\}
\label{coul1w}
\end{align}
where we have introduced vectorial notation $\vec{N}\hat{U}\vec{N}\equiv\sum_{ii'}U_{ii'}N_iN_{i'}$
Here integer $N_i$ denotes a  number of excess electrons on $i$-th grain, $\hat{U}\equiv e^2\hat{C}^{-1}$ is the inverse matrix of capacitances, and the components $Q_i$ of the vector $\vec{Q}$ are the so called ``offset charges'' (not necessarily integers!). It should be noted that each offset charge $Q_i$ can not be identified with certain unique impurity: all impurities that effectively interact with a given grain $i$ contribute to $Q_i$. Vice versa, each impurity may effectively contribute to many different variables $Q_i$.

In the context of VRH the offset charges are usually treated as static random variables, but in this paper we are going to take into account their dynamics. Indeed, $Q_i$ depend on the positions of the charged impurities, that are not absolutely rigid, but can  deviate from their equilibrium places. In the harmonic approximation these deviations are governed by the hamiltonian $\hat{H}_{\rm dev}=\hat{E}_{\rm dev}+\hat{\cal H}_{\rm kin}$,
\begin{align}
\hat{E}_{\rm dev}= \frac{1}{2}(\vec{Q}-\vec{Q}^{(0)})\hat{K}(\vec{Q}-\vec{Q}^{(0)})
\label{deviation1}\\
\hat{\cal H}_{\rm kin}=-\frac{\hbar^2}{2}\partial_{\vec{Q}}\hat{M}^{-1}\partial_{\vec{Q}}
\label{deviation2}
\end{align}
where the positively defined matrix of effective masses $\hat{M}$ is related to the masses of impurities. 
The vector $\vec{Q}^{(0)}$ describes the set of equilibrium values of offset charges for the case of neutral grains: $\vec{N}\equiv 0$.
The matrix $\hat{K}$ in \eqref{deviation1} is related to the stiffness of the system with respect to displacements of the charged impurities. It contains both the ``mechanical'' part (due to deformation of surrounding medium) and the ``electrostatic'' part (the variation of the electrostatic energy of grains \eqref{coul1w} due to the displacement of impurities). The second part depends on the set $\vec{N}$; namely, since $K_{ii'}(\vec{N})$ is linearly related to the second derivative of the energy with respect to the coordinates of impurities, $\hat{K}$ should be a quadratic polynomial in $\vec{N}$:
\begin{align}
K_{ii'}(\vec{N})=K_{ii'}^{(0)}+\sum_jK_{ii'j}^{(1)}N_j+\sum_{jj'}K_{ii'jj'}^{(2)}N_jN_{j'}
\label{deviation3}
\end{align}
The mechanical part contributes only to the first, $N$-independent, term in \eqref{deviation3}. Therefore, in the most natural case, when the mechanical stiffness dominates over the electrostatic one, the $N$-dependence of $\hat{K}$ can be neglected.
Anyway, even if the electrostatic part of $\hat{K}$ is considerable, at low temperatures it seems reasonable to ignore the ``live''  $\vec{N}$-dependence of $\hat{K}$, and replace the function $K_{ii'}(\vec{N})$ by its equilibrium value 
\begin{align}
K_{ii'}\equiv K_{ii'}^{(0)}+\sum_jK_{ii'j}^{(1)}N_j^{\rm (eq)}+\sum_{jj'}K_{ii'jj'}^{(2)}N_j^{\rm (eq)}N_{j'}^{\rm (eq)}
\label{deviation3a}
\end{align}
where $\vec{N}^{\rm (eq)}$ are the charges, that the grains acquire at the equilibrium.

\subsection{The classical ground state \label{The classical ground state}}

Since the effective masses are large, in the leading approximation the kinetic energy term \eqref{deviation2} in the hamiltonian can be neglected, and the ground state of the system corresponds to the minimum of the total potential energy of the system of charges
\begin{align}
E_{\rm ch}(\vec{N},\vec{Q})=E_C(\vec{N},\vec{Q})+E_{\rm dev}(\vec{Q})=\nonumber\\=\frac{1}{2}\left\{(\vec{N}-\vec{Q})\hat{U}(\vec{N}-\vec{Q})-\vec{Q}\hat{U}\vec{Q}\right\}+\nonumber\\+\frac{1}{2}(\vec{Q}-\vec{Q}^{(0)})\hat{K}(\vec{Q}-\vec{Q}^{(0)})
\label{deviation4}
\end{align}
Minimising \eqref{deviation4} with respect to $\vec{Q}$ at fixed $\vec{N}$, we get 
\begin{align}
\vec{Q}_{\min}(\vec{N})=\vec{Q}^{(0)}+\hat{K}^{-1}\hat{U} \vec{N},
\\
\min_{\vec{Q}}E_{\rm ch}(\vec{Q},\vec{N})=\frac12 \vec{N}\hat{\cal U} \vec{N}-\vec{N}\hat{U} \vec{Q}^{(0)}
\label{deviation5}
\end{align}
where the symmetric matrix
\begin{align}
\hat{\cal U}=\hat{U}-\hat{U}\hat{K}^{-1}\hat{U}
\label{deviation5a}
\end{align}
has a meaning of the inverse capacitance matrix, renormalised due to the effects of finite elasticity of the system. 

For the stability of the system's ground state the matrix $\hat{\cal U}$ must be positively defined. A violation of this requirement would mean that we have incorrectly chosen the ground state set $\vec{N}^{\rm (eq)}$, which turned out to be unstable; the system eventually will move to a different state, where the stability will be restored due to nonlinearity of the problem, expressed in the $\hat{K}(\vec{N})$ dependence \eqref{deviation3}.

Further minimisation of \eqref{deviation5} with respect to  $\vec{N}$ gives the equilibrium values of charges as  integers, closest (in a sense, see below) to $\hat{\cal U}^{-1}\hat{U}\vec{Q}^{(0)}$. As a result
\begin{align}
\vec{N}^{\rm (eq)}=\hat{\cal U}^{-1}\hat{U}\vec{Q}^{(0)}+\vec{\gamma},\label{deviation7ll}\\
\vec{Q}^{\rm (eq)}=\vec{Q}^{(0)}+\hat{K}^{-1}\hat{U} \vec{N}^{\rm (eq)}%=\hat{\alpha}\left[1+(1-\hat{\alpha})^{-1}\right]\vec{Q}^{(0)}+\hat{\alpha}\vec{\gamma},
\label{deviation7}
\end{align}
where $\vec{\gamma}$ is the set of ``effective residual offset charges''. Note, that in the case of rigid impurities (when $\hat{K}\to\infty$) $\hat{\cal U}\to\hat{U}$,  and the effective offset charges are reduced to the standard ones: $\vec{\gamma}\to$ noninteger part of $\vec{Q}^{(0)}$

The common restriction, usually imposed on the residual offset charges, reads
\begin{align}
-1/2\leq\gamma_k\leq 1/2
\label{restrict}
\end{align}
Strictly speaking, this is not correct in general case: for symmetric matrix $\hat{\cal U}$ the all-integer-$\vec{N}$ minimum of the energy functional may, in principle, lay quite far from the unrestricted one, so that some components of $\vec{\gamma}$ may be quite large.
Examples are easy to produce (say, a highly anisotropic potential profile with valleys, looking in low-symmetry directions) but all these examples seem to be exotic, if not pathological.  At least we were not able to construct any physically relevant matrix $\hat{\cal U}$ for that the restriction \eqref{restrict} would be violated. Anyway, it is definitely valid for the case which we are going to study in detail below: the screened Coulomb interaction with diagonal matrix $\hat{\cal U}$. Therefore, in what follows we will consider the restriction \eqref{restrict} granted.

\subsection{Low-energy hamiltonian \label{Low-energy hamiltonian}}
 
At low temperatures both the grains' charges $\vec{N}$ and the offset charges $\vec{Q}$ only slightly deviate from the equilibrium values, so one can write
\begin{align}
\vec{N}=\vec{N}^{\rm (eq)}+\vec{n},\quad
\vec{Q}=\vec{Q}^{\rm (eq)}+\vec{q}
\label{deviation7w}
\end{align}
where $n_k$ may take values $(-1, 0, 1)$.
Now we are prepared to rewrite the hamiltonian in terms of deviations $\vec{n}$ and $\vec{q}$. Substituting \eqref{deviation7w} into \eqref{deviation4} and omitting the terms, that do not contain deviations, we get
\begin{align}
E_{\rm ch}
=\frac{1}{2}\left\{(\vec{n}-\vec{q}\;)\hat{U}(\vec{n}-\vec{q})+\vec{q}\;(\hat{K}-\hat{U})\vec{q}\right\}+\vec{n}\;\hat{\cal U}\vec{\gamma}
\label{deviation4g1}
\end{align}
%and the Hamiltonian
% \begin{align}
%\hat{\cal H}_{\rm ch}=-\frac{\hbar^2}{2}\partial_{\vec{q}}\hat{M}^{-1}\partial_{\vec{q}}+E_{\rm ch}(\vec{n},\vec{q})
%\label{deviation2y}
%\end{align}

\subsection{Thermodynamic excitation energies \label{thermodynamic}}

\begin{figure}[ht]
\includegraphics[width=1\columnwidth]{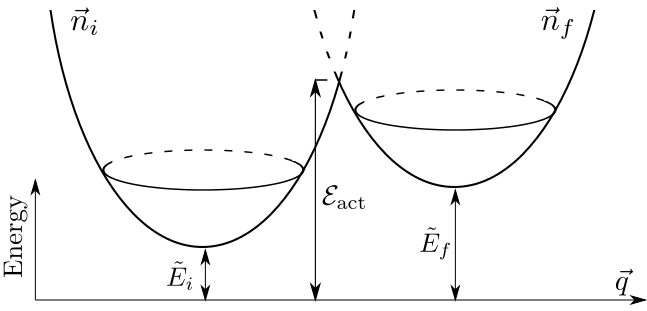}
\caption{Energy paraboloids in the $\vec{q}$-space. The paraboloids are indexed by vector $\vec{n}$. Zero-energy level corresponds to the classical ground state. The activation energy ${\cal E}_{\rm act}$ for transition $\vec{n}_i\to\vec{n}_f$ is given by \eqref{deviation12wc}. } 
\label{paraboloid}
\end{figure}

Besides the ground state, the low-lying excited states are of great importance for the transport properties of the system. For the state $\vec{n}$ the thermodynamic (i.e., minimized with respect to $\vec{q}$) excitation energy is
\begin{align}
\tilde{E}(\vec{n})\equiv \min_{\vec{q}}E_{\rm ch}(\vec{n},\vec{q})=
\frac{1}{2}\vec{n}\;\hat{\cal U}(\vec{n}+ 2\vec{\gamma})
\label{thermodynamicc}
\end{align}
Different branches of spectrum \eqref{deviation4g1} can be visualized as multi-dimensional paraboloids in the $\vec{q}$-space. The paraboloids are indexed by vector $\vec{n}$. Excitation energy $\tilde{E}(\vec{n})$ is nothing else, but the energetic distance between the bottom of corresponding paraboloid and the global ground state energy, see Fig.\ref{paraboloid}. For brevity we have denoted $\tilde{E}_i\equiv\tilde{E}(\vec{n}_i)$ and $\tilde{E}_f\equiv\tilde{E}(\vec{n}_f)$.

\subsubsection{Single-particle excitations \label{Single-particle excitations}}

Below we will obtain some crucial characteristics of single-particle excited states, in which only one entry $n_k$ in $\vec{n}$ is nonzero, while all other entries are zeroes (namely, $\vec{n}=\pm\vec{z}_{k}$ where $[\vec{z}_{k}]_m=\delta_{km}$).

The thermodynamic excitation energies 
\begin{align}
\tilde{E}_k^{(\pm)}=\frac{1}{2}\vec{z}_{k}\:\hat{\cal U}(\vec{z}_{k}\pm 2\vec{\gamma})
\label{deviation8}
\end{align}
correspond to variation of the ``relaxed''  energy due to creation (annihilation) of one electron at grain $k$. By definition, the inequality $\tilde{E}_k^{(\pm)}\ge 0$ should hold for all $k$, which is ensured by \eqref{restrict}. 
In granular systems it is convenient to introduce
\begin{align}
\varepsilon_k=\left\{\begin{aligned} \tilde{E}_k^{(+)},\qquad &\mbox{for
$\vec{z}_{k}\hat{\cal U}\vec{\gamma}<0$}\\
-\tilde{E}_k^{(-)},\qquad & \mbox{for $\vec{z}_{k}\hat{\cal U}\vec{\gamma}>0$}
\end{aligned}
\right.
\label{result1}
\end{align}
that has the meaning of the energy of ``charged ground state'', counted from the global ground state of the grain. 
Density of such states is sometimes called the density of ground states (DOGS) in the literature -- see \cite{ZhangShklovskii,nanocrystals} for more details. 

As we will see soon, another useful combination that enters the activation exponent of the conductance between two distant grains $l$ (by agreement  ``left'') and $r$ (by agreement  ``right'') is
\begin{align}
\varepsilon_{lr}\equiv\frac12\left\{|\varepsilon_l-\varepsilon_r|+|\varepsilon_l|+|\varepsilon_r|\right\}
\label{eps if}
\end{align}
Note that it has a conventional analogue in the standard hopping conductivity theory \cite{ES_book}.

In the following we will assume that electronic transitions proceed very fast, so that the slow variables $\vec{q}$ do not have time to change during the process -- the Franck-Condon principle (see more discussion on this topic in Section \ref{FC principle}). Thus we introduce additional Franck-Condon excitation energies
\begin{align}
    E_k^{(\pm)}(\vec{q})\equiv E_{\rm ch}(\pm\vec{z}_{k},\vec{q})-E_{\rm ch}(0,\vec{q})=\nonumber\\=\tilde{E}_k^{(\pm)}+\frac12\vec{z}_{k}\hat{U}\hat{K}^{-1}\hat{U}\vec{z}_{k}\mp\vec{z}_{k}\hat{U}\vec{q}
\label{deviation9}
\end{align}

\subsubsection{Two-particle excitations \label{Two-particle excitations}}

\begin{figure}[ht]
\includegraphics[width=0.8\linewidth]{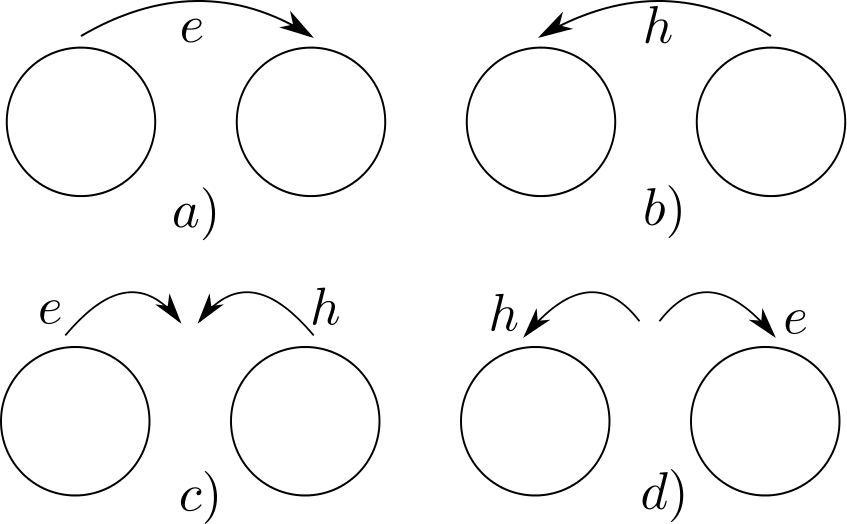}
\caption{Four possible ways of the charge $-e$ transfer from grain $l$ to grain $r$.}
\label{processes}
\end{figure}

There are four classes of possible two-particle processes, corresponding to an act of the charge $-e$ transfer from grain $l$  to grain $r$, see Fig.  \ref{processes}.  
\begin{enumerate}[a)]
\item ($++$ process): Transfer of an electron-type single-particle excitation  from $l$ to $r$: $(\vec{n}=\vec{z}_{l})\longrightarrow (\vec{n}=\vec{z}_{r})$;
\item ($--$ process): Transfer of a hole-type single-particle excitation  from $r$ to $l$: $(\vec{n}=-\vec{z}_{r})\longrightarrow (\vec{n}=-\vec{z}_{l})$;
\item ($+-$ process): Annihilation a two-particle excitation, consisting of an electron-type excitation in grain $l$ and a hole-type one in grain $r$: $(\vec{n}=\vec{z}_{l}-\vec{z}_{r})\longrightarrow (\vec{n}=0)$;
\item ($-+$ process): Creation of a two-particle excitation, consisting of an electron-type excitation in grain $r$ and a hole-type one in grain $l$: $(\vec{n}=0)\longrightarrow (\vec{n}=\vec{z}_{r}-\vec{z}_{l})$;
\end{enumerate}

While in the first two processes only single-particle excitations are involved, in the third and fourth ones the two-particle complexes (intergrain electron-hole pairs) are created or annihilated. Because of generally long-range character of interaction matrix $\hat{\cal U}$ the components of two-particle excitations interact with each other, so that their energies generally are not additive. Thermodynamic excitation energies for the two-particle excitations can be obtained from \eqref{thermodynamicc}:
\begin{align}
\tilde{E}_{lr}^{(\pm\pm)}=\tilde{E}_{l}^{(\pm)}+\tilde{E}_{r}^{(\pm)}\\
\tilde{E}_{lr}^{(\pm\mp)}=\tilde{E}_{l}^{(\pm)}+\tilde{E}_{r}^{(\mp)}-\vec{z}_{l}\hat{\cal U}\vec{z}_{r}
\label{deviation8nn}
\end{align}
and the Franck-Condon energies are
\begin{align}
E_{lr}^{(\pm\pm)}(\vec{q})\equiv E_{\rm ch}(\pm\vec{z}_{l}\pm\vec{z}_{r},\vec{q})-E_{\rm ch}(0,\vec{q})=
\nonumber\\=E_{l}^{(\pm)}+E_{r}^{(\pm)} \\
E_{lr}^{(\pm\mp)}(\vec{q})\equiv E_{\rm ch}(\pm\vec{z}_{l}\mp\vec{z}_{r},\vec{q})-E_{\rm ch}(0,\vec{q})=
\nonumber\\=E_{l}^{(\pm)}+E_{r}^{(\mp)}-\vec{z}_{l}\hat{U}\hat{K}^{-1}\hat{U}\vec{z}_{r}
\label{deviation8nna}
\end{align}

We will see in Section \ref{Short range interaction case} that the interaction parts in \eqref{deviation8nn} and \eqref{deviation8nna} vanish in the case of short-range interaction. 

%\begin{align}
%\varepsilon_{lr}^{\rm (int)}=\vec{z}_{l}\hat{\cal U}\vec{z}_{r}
%\end{align}

\subsubsection{Potential barrier between resonant grains and activation energy \label{Thermodynamic gap}}

A very important role in low temperature physics is played by the resonant grains, for which either $\tilde{E}_k^{(+)}$ or $\tilde{E}_k^{(-)}$ is anomalously close to zero. The transition between such resonant states $\vec{n}_i=\vec{z}_i$ and $\vec{n}_f=\vec{z}_f$ requires, however, a considerable change of the surrounding (i.e., the vector $\vec{q}$), which can only be done continuously. In the course of this change the potential energy of the system also changes -- first increases, then decreases, so that the system has to overcome the potential barrier. This can be accomplished either by means of activation over the barrier, or by tunneling. For both processes the height of the barrier $W$ is crucial. To find it we should minimise the energy $E_{\rm ch}(\vec{n}_i,\vec{q})$ over $\vec{q}$ with additional condition $E_{\rm ch}(\vec{n}_i,\vec{q})=E_{\rm ch}(\vec{n}_f,\vec{q})$, implying that the states are resonant. The result is
\begin{align}
W=\frac{1}{8}(\vec{n}_{f}-\vec{n}_{i})\hat{U}\hat{K}^{-1}\hat{U}(\vec{n}_{f}-\vec{n}_{i})
\label{deviation12w}
\end{align}
Note that $W$ depends only on the difference $\vec{n}_f-\vec{n}_i$ between the final and initial states. In particular, it is the same for all kinds of processes described in Section \ref{Two-particle excitations}.

We can also find the activation energy ${\cal E}_{\rm act}$ for a transition between arbitrary (not necessarily resonant) states $\vec{n}_i$ and $\vec{n}_f$. 
It can be defined as the lowest point of interception of two paraboloids (see Fig. \ref{paraboloid}):
\begin{align}
{\cal E}_{\rm act}=\frac{\tilde{E}_i+\tilde{E}_f}{2}+W+\frac{(\tilde{E}_f-~\tilde{E}_i)^2}{16W}
\label{deviation12wc}
\end{align}
where $W$ is defined by \eqref{deviation12w}. If the states are resonant, we have ${\cal E}_{\rm act}=W$.

\section{Short range interaction model \label{Short range interaction case}}

Those granular system, where the long range part of interaction is screened (say, because of the presence of metallic gate, or due to residual conductivity of the insulating matrix) can be roughly described by the simplest model of ``short range Coulomb interaction'' (see, e.g.,  \cite{NazarovBlanter}). In this model we assume the matrices $\hat{U}$ and $\hat{K}$ to be diagonal
\begin{align}
\hat{U}=
\begin{pmatrix}
\ddots&0&0&0\\
0&U_i&0&0\\
0&0&U_{i+1}&0\\
0&0&0&\ddots
\end{pmatrix}\quad
\hat{K}=
\begin{pmatrix}
\ddots&0&0&0\\
0&K_i&0&0\\
0&0&K_{i+1}&0\\
0&0&0&\ddots
\end{pmatrix}
\nonumber
\end{align}
but their diagonal entries in general are not identical, since different grains have different capacitances etc. The electrostatic energy \eqref{deviation4g1} can be rewritten in a simple way $E_{\rm ch}=~\sum_iE_i(n_i,q_i)$, where
\begin{align}
E_i(n,q)
=E^{c}_i\left\{n^2+2n\left[\gamma_i(1-\alpha_i)-q\right]+\frac{q^2}{\alpha_i}\right\}
\label{diag2a}
\end{align}
and each grain is characterised by three constants: 
\begin{itemize}
\item The standard charging energy
$
E^{c}_i\equiv U_i/2=e^2/2C_i
$
with $C_i$ being the capacitance of the grain $i$,
\item The ``polaronic coupling constant''
$
\alpha_i\equiv U_i/K_i
$.
The inequality $\alpha_i<1$ is the stability condition. 
\item The random ``effective offset charge'' $\gamma_i$, distributed in the interval $[-1/2,1/2]$.
\end{itemize}

The thermodynamic and Franck-Condon excitation energies  for this model are
\begin{align}
\tilde{E}_i^{(\pm)}=E^{c}_i(1-\alpha_i)(1\pm 2\gamma_i),\\
E_i^{(\pm)}(q)=\tilde{E}_i^{\pm}+4W_i \mp 2E^{c}_iq_i,\\
W_i=\alpha_iE^{c}_i/4, \quad W=W_l+W_r
\label{deviation83}
\end{align}

\section{Transition rate\label{Transition rates}}

As we know from \cite{iosel2005}, the main features of variable-range hopping in granular systems can be revealed  already in the simplest model of short range Coulomb interaction, described in Section \ref{Short range interaction case}, so in this paper we restrict our consideration to this model.

\subsection{Franck-Condon principle \label{FC principle}}

We will also ignore the quantum aspects of the offset  charges dynamics (i.e., put $M\to\infty$). In particular, the last assumption means that we can use the Franck-Condon principle in the calculation of transition rates. According to this principle, the set $\vec{q}$ remains unchanged on the cotunneling time scale $\sim E_c^{-1}$. We should calculate the transition rates $w_{ii'}(\vec{q}\;)$ between relevant grains at fixed configuration of $\vec{q}$ and only afterwards perform the thermodynamic averaging of the result with respect to $\vec{q}$. The cotunneling time $\tau_{\rm cotun}$ can be roughly estimated as the inverse scale of energetic denominators appearing in the perturbation theory
\begin{align}
\tau_{\rm cotun}\sim E_c^{-1}
\label{diag1ww}
\end{align}
On the other hand, characteristic time scale $\tau_q$ for the dynamics of $\vec{q}$ is nothing but the inverse frequency of the impurity oscillations $\omega_{\rm imp}$, which apparently should be of the order of phonon frequencies $\omega_{\rm ph}$. As a result, the applicability criterion for the Franck-Condon principle $\tau_{\rm cotun}\ll\tau_q$ becomes
\begin{align}
 \omega_{\rm imp}\ll E_c
\label{diag1ww1}
\end{align}
Note that in typical granular systems the rough estimate $\sim$ 100 K holds for $E_c$ as well as for $\omega_{\rm imp}$. Thus, the condition \eqref{diag1ww1} can not be considered automarically fulfilled, the opposite case is well probable. However, as we already noted, we are going to neglect the kinetic term for $\vec{q}$ in the hamiltonian, which means that $\vec{q}$ can overcome potential barriers only by activation, and not by tunneling. This implies even more stringent condition 
\begin{align}
 \omega_{\rm imp}\ll T\ll E_c
\label{diag1ww2}
\end{align}
that we will consider granted throughout this paper.

\subsection{Hamiltonian and some qualitative considerations}

Under the above approximations, the hamiltonian of the system may be written in the form
\begin{align}
\hat{\cal H}=\sum_i\left\{E_i(n_i,q_i)+\hat{\cal H}_i^{(0)}\right\}+\sum_{\langle ii'\rangle}\hat{\cal H}_{\rm tun}^{(ii')}
\label{diag1}
\end{align}
where $E_i(n_i,q_i)$ is given by expression \eqref{diag2a}, and 
\begin{align}
\hat{\cal H}_i^{(0)}=\sum_{\lambda_i,\sigma}\epsilon_{\lambda_i}\hat{a}^\dagger_{\lambda_i\sigma}\hat{a}_{\lambda_i\sigma}
\label{diag3}
\end{align}
is the hamiltonian of electrons within the $i$-th grain. The index $\lambda_i$ denotes electronic eigenstates with eigenenergies $\epsilon_{\lambda_i}$, which are supposed to be $\sigma$-independent ($\sigma$ is a spin projection). The level spacing $\delta_i$ for electrons at the Fermi level in grain $i$ is small: $\delta_i\ll E_i^{c}$.

The tunneling Hamiltonian
\begin{align}
\hat{\cal H}_{\rm tun}^{(ii')}=\sum_{\lambda_i,\lambda_{i'},\sigma}t_{\lambda_i\lambda_{i'}}\hat{a}^\dagger_{\lambda_i\sigma}\hat{a}_{\lambda_{i'}\sigma}
\label{diag3a}
\end{align}
describes the hops of electrons between neighbouring grains $\langle ii'\rangle$. We are interested in the case, when the typical dimensionless intergrain conductances
\begin{align}
g_{ii'}\equiv |t_{ii'}|^2/\delta_i\delta_{i'}\ll 1
\label{diag3akj}
\end{align}
 are small, and the tunneling hamiltonian may be treated perturbatively; the necessary order of the perturbation theory, however, appears to be high -- the lower the temperature, the higher the order! 
 
The principal idea of any VRH-type calculation is the famous observation of Mott \cite{mott0}: at low temperatures hopping electrons prefer to visit only resonant sites (in our case for sites stand the grains), where their energies are confined to  a narrow strip of width $\varepsilon$ near the Fermi energy. Decreasing $\varepsilon$  loosens the factor $\exp\{-\varepsilon/T\}$ that suppresses the conductivity due to small number of available excitations. On the other hand the resonant sites are rare (the typical distance $r(\varepsilon)$ between them grows with decreasing $\varepsilon$), so that the overlap of corresponding wave-functions $I$ is small: $I\propto\exp\{-2r(\varepsilon)/\xi\}$ and this small factor becomes still smaller with decreasing $\varepsilon$. Thus, one has to find a compromise  between the two exponentially small factors, that results in certain optimal $\varepsilon_{\rm opt}(T)$ and the conductivity $\sigma\propto\exp\{-\varepsilon_{\rm opt}(T)/T\}$. In the presence of polaronic effect the above calculation scenario is somewhat modified, but the main idea remains the same.

While in case of single electron tunneling evaluation of the overlap exponential factor $I$ is straightforward, and $\xi$ is simply related to the decrement of electronic wave-function, in the case of metallic grains the origin of the exponential dependence and explicit form of $\xi$ is more sophisticated. In this section we will explore this problem, incorporating the additional physics that arises from the effects of the stray-charges flexibility.

\subsection{Amplitudes of multiple cotunneling: perturbation theory  \label{Perturbation theory: multiple cotunneling}}

Thus, we are interested in the amplitude of electronic transition between two distant resonant grains $l$ (left) and $r$ (right). Unless there is a tunnelling term in the Hamiltonian, the occupation numbers $\vec{n}$ are preserved. The tunneling term $\hat{\cal H}_{\rm tun}$ allows hops of electrons between neighbouring grains; to accomplish a hop between distant grains, $\hat{\cal H}_{\rm tun}$ has to be applied in some $({\cal N}+1)$-st order of perturbation theory, where ${\cal N}$ is a number of intermediate grains, constituting a continuous chain between $l$ and $r$. A multi-particle process, described by this perturbational approach, is generally known as cotunneling. It was introduced in  \cite{Nazarov1990} and applied to the theory of transport in quantum dots  \cite{NazarovBlanter,Glazman} and granular metals \cite{iosel2005,bel-hopping2}.

\begin{figure}[ht]
\includegraphics[width=1\columnwidth]{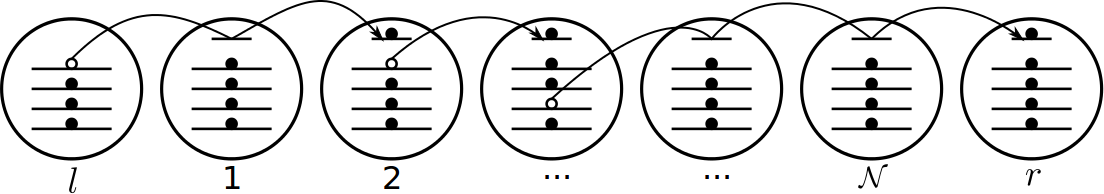}
\caption{Cotunneling process from grain $l$ to grain $r$. } \label{cotun}
\end{figure}

Let's consider the transfer of an electron via virtual states on ${\cal N}$ intermediate grains, $l\equiv0, 1,\ldots ,{\cal N},{\cal N}+1\equiv r$ being the path of adjacent grains, starting at the initial grain $l$ and terminating at the final grain $r$, so that $k$ is a neighbour of $k-1$ and $k+1$ (see Fig.\ref{cotun}). In principle, one should sum over all possible paths connecting $l$ and $r$, but in the case of small tunnelling amplitudes $t$ we can expect the sum to be dominated by shortest paths -- those, with minimal possible number ${\cal N}$ of intermediate grains. Moreover, in most situations only one particular path will be important. The amplitude of such multiple cotunneling event is given by a proper matrix element of 
\begin{align}
A_{\{h_k,e_k\}}(\vec{q})=(-i)^{{\cal N}+1}\nonumber\\\times\int {\rm T}\left\{\hat{S}\hat{{\cal H}}_{\rm tun}(t_{N+1})...{\hat {\cal H}}_{\rm tun}(t_1)\right\}\prod\limits_{k=1}^{{\cal N}+1}dt_k
\label{amp}
\end{align}
calculated at given static set of deviations $\vec{q}$. 
               
The amplitude $A_{\{h_k,e_k\}}$ describes the process, at the end of which a hole with the set of quantum numbers $h_0$ is created in grain $l$ and an electron with the set $e_{{\cal N}+1}$ is created in grain $r$; generally speaking, each of ${\cal N}$ intermediate grains $k=1,\ldots ,{\cal N}$ acquires an electron-hole pair with quantum numbers $\{e_k,h_k\}$ (ineleastic cotunneling). However, it is possible to have $e_k=h_k$ for certain $k$'s, then no electron-hole pairs are created in the corresponding grains (elastic cotunneling). Let us denote the set of such $k$'s as $m_1,m_2,\ldots, m_{\cal M}$. If this set is empty (${\cal M}=~0$), one speaks about purely inelastic multiple cotunneling; if it includes all the intermediate grains (${\cal M}={\cal N}$), we deal with purely elastic multiple cotunneling.

Consider the set of indices $\left\{\alpha_1\beta_1;...;\alpha_{{\cal N}+1}\beta_{{\cal N}+1}\right\}$, which describes certain sequence of individual tunnelings between pairs of neighbouring grains in the chain (in the $k$-th entry a tunnelling event occurs from the state $\alpha_k$ to the state $\beta_k$). Such a set is some permutation of ``the natural sequence'' $\left\{h_0e_1;h_1e_2;...;h_{\cal N}e_{{\cal N}+1}\right\}$, corresponding to progressive motion of electron from the left end of the chain to its right end. All the permuted sets contribute to the amplitude $A_{\{h_k,e_k\}}$ alongside with the natural one.  From \eqref{amp} we get %it follows that the amplitude per unit of time {\bf (???? usually it is the probability, that is proportional to time, not the amplitude!)} is
\begin{widetext}
\begin{align}
\tilde{A}_{\{h_k,e_k\}}=(-i)^{{\cal N}+1}\prod_{k=1}^{{\cal N}+1}t_{e_kh_{k-1}}\int\prod\limits_{k=1}^{{\cal N}+1}d\tau_k
\left\langle\left|
{\rm T}\left\{\hat{\psi}^\dagger_{e_{{\cal N}+1}}(\tau_{{\cal N}}+...+\tau_1)\hat{\psi}_{h_{{\cal N}}}(\tau_{{\cal N}}+...+\tau_1)\cdots
\hat{\psi}^\dagger_{e_1}(0)\hat{\psi}_{h_0}(0)\right\} \right|\right\rangle
\label{part1}
\end{align}
\end{widetext}
where we have used the shortcut notation $|\rangle$ for initial and $\langle|$ for final state. The tunneling matrix element $t_{e_kh_{k-1}}$ describes the transition of electron from the state $h$ in the $k-1$-th grain to the state $e$ in the $k$-th grain.

 Within the short-range interaction model we only need to time-order the $\hat{\psi}$-operators acting in each single-grain subspace:
\begin{align}
-i\left\langle\left|{\rm T}\left\{\psi_{h_k}(\tau_1+\cdots
+\tau_{k})\psi^+_{e_k}(\tau_1+\cdots
+\tau_{k-1})\right\}\right|\right\rangle
\end{align}
As a result, we obtain \eqref{inelastic} for inelastic grains and \eqref{elastic} for elastic ones:
\begin{widetext}
\begin{align}
-in_{h_k}[1-n_{e_k}]\left(\theta(\tau_k)e^{-i(\epsilon_{h_k}+E_k^{(+)})\tau_k-i(\tau_1+\cdots
+\tau_{k-1})(\epsilon_{h_k}-\epsilon_{e_k})}
-\theta(-\tau_k)e^{-i(\epsilon_{h_k}-E_k^{(-)})\tau_k-i(\tau_1+\cdots
+\tau_{k-1})(\epsilon_{h_k}-\epsilon_{e_k})}\right)
\label{inelastic}
\\
-i(1-n_{e_m})\theta(\tau_m)e^{-i(\epsilon_{e_m}+E_k^{(+)})\tau_k}+in_{e_m}\theta(-\tau_m)e^{-i(\epsilon_{e_m}-E_k^{(-)})\tau_k}
\label{elastic}
\end{align}
\end{widetext}

Here $n_s$ is the fermionic occupation number for state~$s$ (not to be mixed with the occupation numbers of grains!). The two terms in \eqref{inelastic}, \eqref{elastic} correspond to different sequences of tunnelings: in one case the hole is created first and the electron afterwards, and in another case the order is reversed. Let us now collect all the factors containing $\tau_k$ for each $k\in1...{\cal N}$ in \eqref{part1} and integrate them out. Then we eventually arrive at
\begin{align}
\tilde{A}_{\{h_k,e_k\}}=-in_{h_0}[1-n_{e_{{\cal N}+1}}]\prod_{k=1}^{{\cal N}+1}t_{e_kh_{k-1}}\prod_{k=1}^{{\cal N}}B_k
\label{amp2}
\end{align}

The factors $B_k$ for inelastic and elastic grains respectively are
\begin{align}
B_k(\vec{q})=-n_{h_k}[1-n_{e_k}]\nonumber\\\times\left\{\frac{1}{\epsilon_{h_k}+\Delta_k+E_k^{(+)}(q_k)}+
\frac{1}{\epsilon_{h_k}+\Delta_k-E_k^{(-)}(q_k)}\right\}, \nonumber\\
B_m(\vec{q})=\nonumber\\=-\left\{\frac{1-n_{e_m}}{\epsilon_{e_m}+\Delta_m+E_m^{(+)}(q_m)}+\frac{n_{e_m}}{\epsilon_{e_m}+\Delta_m-E_m^{(-)}(q_m)}\right\}
\label{Bs}
\end{align}
where we have introduced
\begin{align}
\Delta_k\equiv-\epsilon_{e_{{\cal N}+1}}-E_r^{(+)}(q_r)+\sum_{s=k+1}^{{\cal N}}(\epsilon_{h_s}-\epsilon_{e_s})
\end{align}

Note that the explicit summation over all tunnelling sequences, which we have just performed, is only possible in the short-range interaction model.

According to Fermi golden rule the transition rate is
\begin{align}
w_{lr}(\vec{q})=2\pi\sum\limits_{\{h_k,e_k\}}|\tilde{A}_{\{h_k,e_k\}}|^2\nonumber\\\times\delta\big(\epsilon_{h_0}-\epsilon_{e_{N+1}}+\sum\limits_{s=1}^{\cal N}(\epsilon_{h_s}-\epsilon_{e_s})-\Delta\big)
\label{trans}
\end{align}
where the Franck-Condon energy difference
\begin{align}
\Delta\equiv\Delta(\vec{q}\:)=E_{\rm ch}(\vec{n}_f,\vec{q}\:)-E_{\rm ch}(\vec{n}_i,\vec{q}\:)
\end{align}
is the difference between electrostatic energies of final and initial states. We should also define the thermodynamic energy difference
\begin{align}
\tilde{\Delta}=\tilde{E}_f-\tilde{E}_i
\end{align}
which is the energy distance between two local minima corresponding to initial $\vec{n}_i$ and final $\vec{n}_f$ states.

Note that it is impossible to distinguish the final states of the systems with the same sets of inelastic quantum numbers and different sets of elastic quantum numbers. For this reason we should add the amplitudes of such processed rather than probabilities. However, because of violent sign fluctuations of tunneling elements we will neglect such interference terms, which is implied in \eqref{trans}.

\section{Averaging of the transition rate \label{Averaging of the transition rate}}

The transition rate \eqref{trans} depends on  both dynamic random variables $\vec{q}$ and static ones -- $\vec{\gamma}, \vec{\alpha}$ etc. There is an important difference between these two groups of variables: we should perform thermodynamic averaging of \eqref{trans} over $\vec{q}$, while the frozen static variables do not imply averaging, so that, in principle, they remain ``live'' and characterise a specific surrounding of a particular chain.   However, because of typically large number ${\cal N}$ of grains in the chain the static disorder is partly self-averaged and washed out. As we will see, only the dependence on few static random variables (mainly those, characterising the terminal grains $l$ and $r$) remains live.

\subsection{Thermodynamic averaging \label{Thermodynamic averaging}}

Gibbs thermodynamic averaging over $\vec{q}$ has the form 
\begin{align}
\langle(...)\rangle_{\vec{q}}=\frac{1}{Z_c}\sum\limits_{\vec{n}}\int  (...) \exp\left\{-\frac{E_{\rm ch}(\vec{n},\vec{q})}{T}\right\}d\vec{q}
\end{align}
where $Z_c$ is the partition function:
\begin{align}
Z_c=\left[\det\frac{\hat{K}}{2\pi T}\right]^{-1/2}   \sum\limits_{\vec{n}}\exp\left\{-\tilde{E}(\vec{n})/T\right\}
\end{align}
In averaging over electronic states we will assume an equilibrium noncorrelated distribution, so that  $\langle n_{s_k}\rangle=f_F(\epsilon_{s_k})$ and $\langle n_{s_k}n_{s_l}\rangle=\langle n_{s_k}\rangle\langle n_{s_l}\rangle$, with $f_F$ being the Fermi function.

Further simplification can be made if we note that the characteristic value of inelastic energies $\epsilon_{h_l}$, $\epsilon_{e_l}$ is controlled by the combination of Fermi-functions and delta-function in \eqref{trans}. We can see that $\epsilon_{\rm inel}\sim\Delta/L$, where $L={\cal N}-{\cal M}$ is the number of inelastic grains in the string. This means that typically $\epsilon_{\rm inel}\ll E_{c}$ . The same is also true for $\Delta_k$. Therefore, these quantities in the denominators in \eqref{Bs} can be neglected compared to $E_k^{(\pm)}$. However, this is not true for the elastic energies $\epsilon_m$, since they don't enter the delta-function. These energies are of order $E_c$ and should be kept in the denominators. For the same reason (as $\epsilon_{h_m}\sim E_c\gg T$) we can substitute $1-\theta(\epsilon_{h_m})$ instead of $f_F(\epsilon_{h_m})$.

The summation over spin projections should be performed only for inelastic grains. It is straightforward and yields the factor $2^{L+1}$. 

Also, we can substitute $t_{e_kh_{k-1}}$ by its ``coarse grained'' value at the Fermi level $t_{k,k-1}$ (i.e., averaged over an interval of energies, large compared to the level spacing $\delta$, but small compared to any other relevant scale). It allows for replacement of the summation over electronic states by integration $\sum_{\epsilon_{s}}\rightarrow \int d\epsilon_s/\delta_k$, which immediately allows to integrate out the elastic energies, and we arrive at:
\begin{widetext}
\begin{align}
\overline{w}_{if}=2\pi\left\langle2^{L+1}\prod\limits_{k=1}^{{\cal N}+1}|t_{k,k-1}|^2 \sum\limits_{\{h_l,e_{l}\}}f_F(\epsilon_{h_0})[1-f_F(\epsilon_{e_{{\cal N}+1}})]
\prod\limits_{k, \text{\rm inelastic}}f_F(\epsilon_{h_k})[1-f_F(\epsilon_{e_{k}})]\left[\frac{1}{E_k^{(+)}}-
\frac{1}{E_k^{(-)}}\right]^2
\times\right.\nonumber\\ \times
\prod\limits_{m,\text{\rm elastic}}\left[\frac{1}{E_m^{(+)}}+ \frac{1}{E_m^{(-)}}\right]\left.
\delta\left(\epsilon_{h_0}-\epsilon_{e_{{\cal N}+1}}+
\sum_{l=1}^{\cal N}(\epsilon_{h_l}-\epsilon_{e_l})-\Delta\right)\right.\Bigg\rangle_{\vec{q}}
\label{large}
\end{align}
\end{widetext}

The variables $q_k$ for intermediate grains enter $\overline{w}_{if}$ only through the  denominators in square brackets of both types (elastic and inelastic), and the result of integration over $dq_k$ in thermal averaging formally diverges at resonances, where either $E^{(-)}_k$ or  $E^{(+)}_k$ goes to zero. This divergency, however, should be cut off at $E^{\pm}_k\sim |t|$, where the perturbation theory ceases to be valid, thus the integration in the vicinity of the resonance gives finite result. So, each integration over $dq_k$ gives three contributions: one from the vicinity of thermal equilibrium $q_k\approx 0$, and the other two from the vicinities of the resonances $q_k^{(\pm)}$, defined by conditions $E^{(\pm)}_k(q_k^{(\pm)})\approx 0$. The relative magnitudes of the resonant contributions are exponentially suppressed $\sim E_c/|t|\exp\{-E_c/T\}$ and thus are negligible in the low-temperature VRH regime. As a result, thermodynamic fluctuations of $q$ at intermediate grains are not relevant, and one can simply neglect them, putting $q_k\approx 0$, so that $E^{(\pm)}_k(q_k)\to E^{(\pm)}_k(0)$ in \eqref{large}.

\subsection{Self-averaging of intermediate grains\label{Self-averaging of intermediate grains}}

To treat the products of large number of random factors, occurring in \eqref{large}, we will apply the CLT self-averaging rule in the form
\begin{align}
\prod _{i=1}^nX_i\approx\exp\left\{n\,\overline{\ln X}\right\},\quad n\gg 1
\label{self}
\end{align}
and hence
\begin{align}
|t_{kk-1}|^2\to t^2\equiv e^{\overline{\ln |t_{kk-1}|^2}}								\nonumber\\ 
\delta_k\to \delta\equiv e^{\overline{\ln\delta_k}}										\nonumber\\
E^{(c)}_i\to E_{c}\equiv e^{\overline{\ln E_k^{(c)}}}									\nonumber\\
\left(\frac{1}{E_m^{(+)}(0)}+ \frac{1}{E_m^{(-)}(0)}\right)\to\frac{A_1}{E_{c}}			\nonumber\\ 
\label{self2}
\end{align}
\begin{align}
\left(\frac{1}{E_k^{(+)}(0)}-\frac{1}{E_k^{(-)}(0)}\right)^2\to\frac{A_2}{E_{c}^2}
\end{align}
where 
\begin{align}
A_1\equiv 2\exp\{-\overline{\ln[1-4(1-\alpha)^2\gamma^2]}\}\nonumber\\
A_2\equiv 4A_1^2\exp\{\overline{\ln[(1-\alpha)\gamma]}\}
\label{self2a}
\end{align}
The factors $A_1$, $A_2$ depend on the specific distribution functions of random parameters $\alpha$ and $\gamma$ and their correlation. They are model-dependent numbers of order unity and we will never refer to their exact values in this paper.

\subsection{Back to thermodynamic averaging\label{Back to thermodynamic averaging } }

Finally, we have to perform the remaining summation over the energies $\epsilon_{e_l},\epsilon_{h_l}$ of the components of electron-hole pairs, created in the inelastic grains, and over the vector $\vec{q}$.

Let us introduce the new temperature scale 
\begin{align}
T_*=\sqrt{(A_1/2A_2)E_c\delta}
\label{croqq}
\end{align}
and more convenient  notations $\epsilon_{l}=\epsilon_{e_l}$, $\epsilon_{{\cal N}+l}=-\epsilon_{h_{l}}$ for all "inelastic" intermediate grains, and $\epsilon_{2{\cal N}+1}=\epsilon_{e_{{\cal N}+1}}$, $\epsilon_{2{\cal N}+2}=-\epsilon_{h_0}$ for terminal grains. Then we can rewrite \eqref{large} as
%\begin{widetext}
\begin{align}
\overline{w}_{if}=4\pi g T_*\left(\frac{A_1g\delta }{E_c}\right)^{\cal N}\sum_{L=0}^{{\cal N}} C_{\cal N}^{L}\left\langle I_L\left[\Delta(q_i,q_f)\right]\right\rangle_{\vec{q}}
\label{xxx}
\end{align}
where 
\begin{align}
I_L(\Delta)=T_*\int\limits_{-\infty}^{\infty} 
\prod_{l=1}^{2L+2} \frac{d\epsilon_{l}}{T_*}[1-f_F(\epsilon_{l})]\delta\left(\sum_{l=1}^{2L+2}\epsilon_{l}+\Delta\right)\label{large2}
\end{align}
%\end{widetext}

The binomial coefficient $C_{\cal N}^L\equiv {\cal N}![({\cal N}-L)!L!]^{-1}$ appears in the formula \eqref{xxx} as a number of possible partitions of the string into elastic and inelastic subsets. Note that in \eqref{large2} we have used the relation \eqref{diag3akj} and expressed the result in the terms of the average dimensionless conductance between adjacent grains $g\equiv (|t|/\delta)^2\ll 1$.

As it was shown in \cite{iosel2005}, in the absence of the polaronic effect (for $W\to 0$) the characteristic scale $\epsilon_{\rm inel}$ for the energies of electron-hole pairs, created in the acts of inelastic cotunneling, is much larger than temperature. It allows for an evaluation of the multiple integral in \eqref{large2}, leading to the result:
\begin{align}
I_L(\Delta)=\frac{(|\Delta|/T_*)^{2L+1}}{(2L+1)!}\exp\left\{-\frac{\Delta+|\Delta|}{2T}\right\}, \quad \frac{\Delta}{L}\gg T
\label{something-k}
\end{align}

A physical meaning of this result is clear: if the number of inelastic grains is $L$ and their total energy is $\Delta$, then the characteristic energy of one electron (or one hole), created in the process is $\epsilon_{\rm inel}\sim\Delta/L$. The phase volume, corresponding to the processes with $2L+2$ particles with energies $\epsilon\sim\epsilon_{\rm inel}$ and finite density of states is then proportional to $\epsilon_{\rm inel}^{2L+2}$, which (with the help of Stirling formula) explains formula \eqref{something-k}.

We will see soon, that in the presence of the polaronic effect (namely, for $\Delta\lesssim W$) $\epsilon_{\rm inel}$ becomes comparable to temperature. Finding $I_L(\Delta)$ for $\epsilon_{\rm inel}\lesssim T$ is a much more difficult problem, which, however, is possible to resolve, using a trick, proposed in \cite{melnikov}. 
 
We should Fourier-transform the $\delta$-function in \eqref{large2} to decouple the integrals over $\epsilon_l$:
\begin{align}
\frac{1}{2\pi}\int e^{-i\Delta t}dt \prod\limits_{l=1}^{2L+2}\int \frac{e^{-it\epsilon_l}}{1+e^{-\epsilon_l/T}}d\epsilon_l
\label{str2}
\end{align}

The trick is to multiply it by $1=\exp(-\frac{\Delta}{2T}-\frac{1}{2T}\sum\epsilon_l)$, which will make the integrals convergent, and they can be easily calculated via residue theory:

\begin{align}
\frac{1}{2\pi}e^{-\Delta/2T}\int e^{-i\Delta t}dt \prod\limits_{l=1}^{2L+2}\int \frac{e^{-it\epsilon_l}}{2\cosh{\frac{\epsilon_l}{2T}}}d\epsilon_l=
\nonumber
\end{align}
\begin{align}
=\frac{1}{2\pi}e^{-\Delta/2T}\int e^{-i\Delta t}dt \left[\frac{\pi T}{\cosh \pi t T}\right]^{2L+2}
\end{align}
As a result,
\begin{align}
I_L(\Delta)=\frac{T_*}{2\pi}e^{-\Delta/2T}\int e^{-i\Delta t}dt \left(\frac{\pi T/T_*}{\cosh \pi t T}\right)^{2L+2}
\label{meln1}
\end{align}
The integral \eqref{meln1} can be evaluated exactly, but it is more convenient first to perform summation over $L$, which  in this representation turns out to be trivial:
\begin{align}
\sum_{L=0}^{{\cal N}} C_{\cal N}^{L}I_L(\Delta)=\frac{T_*}{2\pi}e^{-\Delta/2T}\nonumber\\\times\int e^{-i\Delta t}dt \left[1+\left(\frac{\pi T/T_*}{\cosh \pi t T}\right)^2\right]^{2{\cal N}}\left(\frac{\pi T/T_*}{\cosh \pi t T}\right)^2
\label{meln2}
\end{align}
We are left with the thermodynamic averaging $\langle\exp\{-\Delta(1/2T+it)\}\rangle_{\vec{q}}$, which is reduced to the gaussian integration that can be easily performed. As a result
\begin{widetext}
\begin{align}
\overline{w}\propto\left(\frac{A_1g\delta }{E_c}\right)^{\cal N} \exp\left\{-\frac{\varepsilon_{lr}}{T}-\frac{D^2}{16WT}\right\}
\int \exp\left\{-\frac{W}{T}\left(2tT-i-i\frac{D}{4W}\right)^2\right\}\left[1+\left(\frac{\pi T/T_*}{\cosh\pi tT}\right)^2\right]^N\frac{dt}{\cosh^2\pi tT}
\label{large3}
\end{align}
\end{widetext}
where 
\begin{align}
 D\equiv |\tilde{\Delta}|- 4W, \qquad \tilde{\Delta}=\varepsilon_l-\varepsilon_r
\label{something1ss2}
\end{align}
and we have  omitted all preexponential factors. 
% \tilde{E}_i+\tilde{E}_f=|\varepsilon_l|+|\varepsilon_r|

\section{Case studies: Transition rates at different strength of polaronic effect \label{different strength} }

The general formula \eqref{large3}, in principle, contains the answers for all possible questions concerning different modes of the charge transfer between two particular grains.  However, for understanding of the physical origin of each particular mode and the corresponding $T$-dependences,  it is necessary to consider the limiting cases. In this Section we undertake such a case study. 

To evaluate the integral over $t$ one can use the steepest descent method. The saddle point is located below the lowest pole on the imaginary axis of the complex $t$ plane at $t=i(1-\xi)/2T$, where $\xi$ satisfies the equation
\begin{align}
D+4W\xi=2\pi {\cal N}Ty(\xi)\cot(\pi\xi/2)
\label{saddle1}
\end{align}
and
\begin{align}
y(\xi)\equiv \left[1+\left(\frac{T_*\sin (\pi\xi/2)}{\pi T}\right)^2\right]^{-1}=\frac{\overline{L}}{\cal N}
\label{saddle1x}
\end{align}
determines the typical number $\overline{L}$ of inelastic grains in the string.
In terms of $\xi$ the transition rate can be written as
\begin{align}
\overline{w}\propto\left(\frac{A_1g\delta}{E_c}\right)^{\cal N} \left[1+\left(\frac{\pi T}{T_*\sin (\pi\xi/2)}\right)^2\right]^{\cal N}\nonumber\\\times\exp\left\{\frac{W}{T}\xi^2-\frac{\varepsilon_{if}}{T}+\frac{D}{2T}\xi\right\}
\label{big1}
\end{align}

We have omitted in \eqref{big1} the preexponential factor $\sin^{-2} \pi\xi$, because it becomes essential only at extremely low temperatures in the case of purely elastic cotunneling. This factor may be important for systems where the length of strings ${\cal N}$ is restricted from above (as in small arrays of quantum dots, or single-electron transistors), which is not the case as long the VRH conductivity of a large sample is concerned. 

We will see that the entire range of the parameter $D$ ($-4W<D<\infty$) may be split into three intervals with different types of approximations applicable:
\begin{enumerate}
\item Weak polaronic effect ($D>0$, $D\gg \Delta D$)
\item Strong polaronic effect ($D<0$, $|D|\gg \Delta D$)
\item Narrow transition range ($|D|\lesssim \Delta D$)
\end{enumerate}
The transition range width
\begin{align}
\Delta D=8\sqrt{W \overline{L}T}\lesssim 8\sqrt{W {\cal N}T}\ll 4W
\label{tra1}
\end{align}
The latter inequality is valid because, as we will see in the Section \ref{Conductivity}, $\max\{W,|\tilde{\Delta}|\}/T\gg{\cal N}$ for typical strings of grains that contribute to the conductivity. Below we discuss these three intervals separately.

\subsection{Weak polaronic effect: ``Electron Hopping''\label{Weak polaronic effect: ``Electron Hopping''}}

For $D>0$ the parameter $\xi\ll 1$, so that all trigonometric functions in \eqref{saddle1} and \eqref{saddle1x} can be expanded. Besides that, as long as  $D\gg \Delta D$ one can neglect the second term on the left hand side of \eqref{saddle1}, so that the latter is reduced to the cubic equation
\begin{align}
(T_*/2T)^2\xi^3+\xi=\frac{4{\cal N}T}{D}
\label{saddle1e}
\end{align}
and one can easily express $\xi$ via $y$:
\begin{align}
\xi=\frac{2T}{T_*}\sqrt{(1-y)/y}
\label{rret}
\end{align}
The transition rate then can be rewritten as
\begin{align}
\overline{w}\propto\left(\frac{A_1g\delta}{E_c(1-y)}\right)^{\cal N} e^{\frac{D}{2T}\xi(y)}
\exp\left\{-\frac{\varepsilon_{if}}{T}\right\}
\label{big1w}
\end{align}
Substituting \eqref{rret} to \eqref{saddle1e}, we arrive at the equation
\begin{align}
y^3=(1-y)z^2, \quad \text{\rm where}\quad z=\frac{D}{2T_*N}
\label{sum4a}
\end{align}
which implicitly determines the function $y(z)$.
 Finally, for the transition rate we obtain
\begin{align}
\overline{w}_{if}\sim\left[\frac{A_1g\delta }{E_c}\exp\left\{\Theta_1\left(\frac{D}{2T_*{\cal N}}\right)\right\}\right]^{\cal N}\exp\left\{-\frac{\varepsilon_{if}}{T}\right\}
\label{fin2v}
\end{align}
the function $\Theta_1(z)$ being defined as
\begin{align}
\Theta_1(z)=2y(z)-\ln(1-y(z))\label{sum5a}
\end{align}
The asymptotics of $\Theta_1(z)$ are
\begin{align}
\Theta_1(z)\approx\left\{\begin{aligned}
3z^{2/3},\qquad &\mbox{for
$z\ll 1$}\\
2\ln(ze),\qquad & \mbox{for $z\gg 1$}
\end{aligned}
\right. 
\label{sum5e}
\end{align}
Note, that $\Theta_1(z)$ coincides with $\varphi(z)$ which was described in \cite{iosel2005}, so that for $D>0$ their result coincides with the result of the present paper up to a slight modification $|\tilde{\Delta}|\to D$ in the definition of the argument $z$.

One can easily check, that the parameter $\xi$ is indeed small under conditions $D>0$, $\max\{W,|\tilde{\Delta}|\}\gg \overline{L}T$, no matter whether $T<T_*$ or $T>T_*$.

\subsection{Strong polaronic effect: ``Polaron Hopping''\label{Strong polaronic effect: ``Polaron Hopping''}}

For negative and not very small $D$ we can neglect the right hand side of equation \eqref{saddle1} (it can definitely be done if $|D|\gg\delta D$), and then
\begin{align}
\xi=|D|/4W\label{tyr}
\end{align}
where $\xi$ is not necessarily small. Substituting \eqref{tyr} into \eqref{big1}, we get
\begin{align}
\overline{w}_{lr}\propto\left[\frac{A_1g\delta }{E_c}\exp\left\{\Theta_2\left(\frac{T_*}{\pi T}\sin\frac{\pi|D|}{8W}\right)\right\}\right]^{\cal N}\nonumber\\ \times 
\exp\left\{-\frac{\varepsilon_{lr}}{T}-\frac{D^2}{16WT}\right\}
\label{large3dd}
\end{align}
and also
\begin{align}
y=\frac{1}{1+z^2}, \qquad \text{\rm где }\; z=\frac{T_*}{\pi T}\sin\frac{\pi|D|}{8W}
\label{saddle1xw}
\end{align}
In order to write the transition rate in a compact form, like \eqref{fin2v}, we have introduced the new function $\Theta_2(z)$
\begin{align}
\Theta_2(z)=\ln(1+1/z^2). \label{theta2}
\end{align}

\subsection{Narrow transition range}

Looking at the results of the two preceding subsections, we conclude that $\xi\ll 1$ for all positive $D$ and also for small negative $D$, (such that $|D|\ll 4W$). However, for very small $|D|$ (both positive and negative) the first term on the left hand side of \eqref{saddle1} can not be neglected. The condition of ``very small''  $|D|$ reads $|D|\sim \xi W$, where $\xi$ is given by the solution of \eqref{saddle1e}. It yields $|D|\sim \Delta D$, where $\Delta D$ is given by \eqref{tra1}. To find $\xi$ in the narrow transition range $|D|\lesssim \Delta D$ one would have to solve a quartic equation
\begin{align}
[(T_*/2T)^2\xi^3+\xi]\left[1+\frac{4\xi W}{D}\right]=\frac{4{\cal N}T}{D}
\label{saddle1euu}
\end{align}
To justify the numerical coefficients in \eqref{tra1}, we note that the saddle point equation \eqref{saddle1} for small $\xi$ can be written in the form
\begin{align}
D+4W\xi=\frac{4\overline{L}T}{\xi}
\end{align}
where $\overline{L}\equiv y(\xi){\cal N}$  itself depends on $\xi$.
A formal ``solution'' of this equation is
\begin{align}
\xi=-\frac{D}{8W}\left(1\pm\sqrt{1+\frac{64W\overline{L}T}{D^2}}\right)
\label{sol}
\end{align}
From the result \eqref{sol} immediately follows the estimate for the width of transition range $\Delta D\sim8\sqrt{W\overline{L}T}$. Since  the requirement $\overline{L}\leq \cal N$ is always satisfied, we arrive at the inequality in \eqref{tra1}.

 We will not discuss the transition region in detail, since the corresponding range of random parameters $D,W$ is narrow and does not give any considerable contribution to physical observables.

\section{Physical interpretation \label{physical explanation}}

Activation exponential factors in formulas \eqref{fin2v} and \eqref{large3dd} coincide with the corresponding factors from standard polaron hopping theory. Additional modifications, specific for our problem, arise only in the power-law factors (effective overlap integrals) due to the many-particle nature of the cotunnelling process. 

\subsection{Main exponential dependence}

Activation exponential factors for conventional polarons were obtained previously by other researchers. Still, we would like to present some physical arguments that qualitatively explain the origin of these factors.

\begin{figure}[ht]
\includegraphics[width=1\columnwidth]{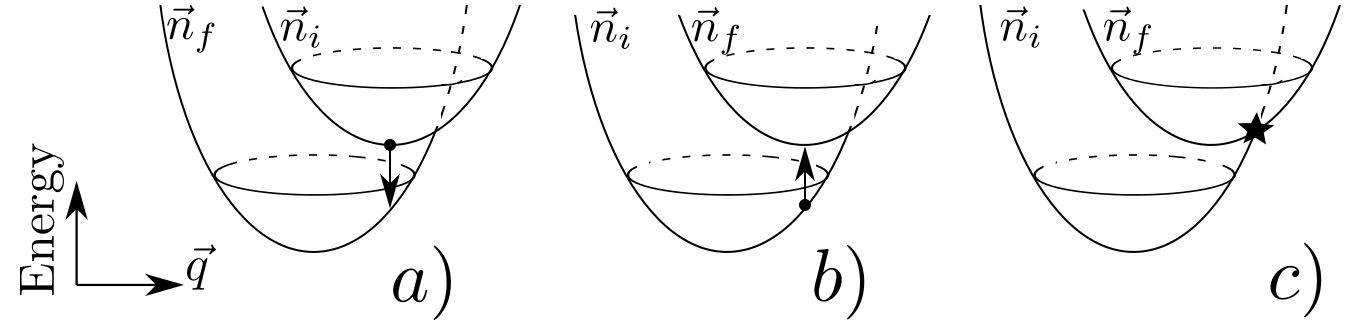}
\caption{Different types of transitions: a) exothermic b) endothermic c) polaron transition.} \label{pic1}
\end{figure}

Let's firstly consider the exothermic transition with $\Delta<0$, i.e. $E_{\rm ch}(\vec{n}_f,\vec{q})<E_{\rm ch}({\vec{n}_i,\vec{q}})$, (see Fig.\ref{pic1} a). Since the total electrostatic energy decreases, the electron-hole pairs will be created in the inelastic intermediate grains to ensure energy conservation. Thus, in this situation with exponential accuracy the probability of transition is just the probability to find the system in state $\vec{n}_i$. Its maximum value is   
\begin{align}
\overline{w}\propto \exp\left\{-\tilde{E}_i/T\right\}
\label{case1}
\end{align}

Now let's consider the endothermic transition with $\Delta>0$, i.e. $E_{\rm ch}(\vec{n}_f,\vec{q})>E_{\rm ch}({\vec{n}_i,\vec{q}})$, (see Fig.\ref{pic1} b). Since the electrostatic energy increases, the electron-hole pairs will be annihilated in some intermediate grains to make up the shortfall. Such pairs are difficult to find at low temperature, which is accounted for by the additional exponential factor $\exp(-\Delta/T)$. The transition rate also contains $\exp\left\{-E_{\rm ch}(\vec{n}_i,\vec{q})/T\right\}$ -- the probability to find the system in state $\vec{n}_i$. Together these two contributions yield $\overline{w}\propto\exp\left\{-E_{\rm ch}(\vec{n}_f,\vec{q})/T\right\}$, and we should take its maximum value: 
\begin{align}
\overline{w}\propto \exp\left\{-\tilde{E}_f/T\right\}
\label{case2}
\end{align}

But what if $\vec{q}$, that delivers the minimum of $E_{\rm ch}(\vec{n}_f,\vec{q})$, violates the requirement $\Delta>0$? The result \eqref{case2} will be incorrect in this case, since there will be no exponential factor $e^{|\Delta|/T}$ for $\Delta<0$! In this case the true minimum of the activation energy should lay on the boundary of the regions with $\Delta>0$ and $\Delta<0$, i.e. at $\Delta=0$, (see Fig.\ref{pic1} c).
\begin{align}
\overline{w}\propto  \exp\left\{-{\rm min}\; E_{\rm ch}(\vec{n}_f,\vec{q})/T\right\}, \quad \vec{q}:\;\Delta=0
\label{str}
\end{align}

Finding the minimum, we obtain
\begin{align}
\overline{w}\propto \exp\left\{-{\cal E}_{\rm act}/T\right\}
\label{case3}
\end{align} 
where ${\cal E}_{\rm act}$ is given by \eqref{deviation12wc}. 

To determine which of the above described solutions gives the largest contribution to the probability of transition, we should simply compare the corresponding exponents. It is easily seen that the contribution of the boundary minimum \eqref{case3} dominates when $|\tilde{E}_i-\tilde{E}_f|<4W$ and explains the main exponential dependence in Polaron Hopping regime \eqref{large3dd}. In the opposite case $|\tilde{E}_i-\tilde{E}_f|>4W$ we have to choose the maximum of \eqref{case1} and \eqref{case2}, which explains the result $\overline{w}\propto\exp\left\{-\varepsilon_{if}/T\right\}$ from \eqref{fin2v}.

\subsection{Energies of electron-hole pairs \label{Energies of electron-hole pairs}}

In the Electron Hopping regime the Franck-Condon transition energy $\Delta$ (that coincides with the aggregate energy of electron-hole pairs involved) doesn't differ much from its ``relaxed'' value $\tilde{\Delta}$. Hence 
\begin{align}
\varepsilon_{\rm inel}\sim\Delta/\overline{L}\approx\tilde{\Delta}/\overline{L}\sim {\cal L}T\gg T
\end{align}
where $\cal L$ is the large logarithmic factor (see later). 

On the contrary, according to the above described physical picture, in Polaron Hopping regime the transition occurs at $\Delta=0$, see \eqref{str}. However, presented in Section \ref{different strength} more careful calculation (which takes into account the $\Delta$-dependence of the power-law factor) shows that $\Delta$ is indeed small, but finite, in contrast with conventional polaron transitions (where it is exactly zero). In fact, $\Delta\sim \overline{L}T$, which means that in the Polaron Hopping regime 
\begin{align}
\varepsilon_{\rm inel}\sim \Delta/\overline{L}\sim T
\end{align}  
Another way to see this is to note that in the presence of polaronic effect the integral \eqref{large3} converges at $t\lesssim 1/T$, which means that the integrals over the inelastic energies \eqref{str2} converge solely due to the denominators, which means at $\varepsilon_{\rm inel}\sim T$.

Thus, we conclude that during the crossover from electron to polaron hopping the characteristic energies of electron-hole pairs decrease from $\varepsilon_{\rm inel}\sim {\cal L}T$ to $\varepsilon_{\rm inel}\sim T\ll {\cal L}T$.

\section{Conductivity: general consideration\label{Conductivity}}

We now finally turn to the calculation of conductivity of a sample of granular metal. According to the general philosophy of the hopping, we introduce the Miller-Abrahams network of conductances $g_{ij}\propto \overline{w}_{ij}$, connecting each pair of grains. As usual, at low temperatures the conductivity is dominated by distant pairs of resonant grains, so that we can use the results of the preceding Section, and represent the conductances in (almost) standard form: 
\begin{align}
g_{ij}\sim\exp\left\{-\frac{{\cal E}_{ij}}{T}-\frac{2|{\bf r}_i-{\bf r}_j|}{a_{ij}}\right\}
\label{fin2vq}
\end{align}
where ${\bf r}_i$, ${\bf r}_j$ are the positions of the centres of the grains and the ``effective decay length'' $a_{ij}$ is given by
\begin{align}
\frac{2a_0}{a_{ij}}=\ln\left(\frac{E_c}{A_1g\delta}\right)-\left\{\begin{aligned}\Theta_1\left(\frac{D_{ij}}{2T_*{\cal N}_{ij}}\right),&\quad \mbox{$D_{ij}>0$}\\
\Theta_2\left(\frac{T_*}{\pi T}\sin\frac{\pi |D_{ij}|}{8W_{ij}}\right),&\quad \mbox{$D_{ij}<0$}\end{aligned}\right.
\label{averaging1bk}
\end{align}
where $a_0$ is the average distance between neighbouring grains, so that $|{\bf r}_i-{\bf r}_j|\equiv a_0{\cal N}_{ij}$.  The functions $\Theta_1$ and $\Theta_2$ are given by \eqref{sum5e} and \eqref{theta2}.

The dependence of $a_{ij}$ on both $r_{ij}$ and characteristics of grains seems to be unusual, however it is a distinct feature of hopping in granular materials. This dependence is only logarithmic and therefore it can be taken into account perturbatively (see later).

The activation energy is standard for a polaronic problem:
\begin{align}
{\cal E}_{ij}=\varepsilon_{ij}+\left\{\begin{aligned}0,&\quad \mbox{($D_{ij}>0$)}\\
\frac{D_{ij}^2}{16W_{ij}},&\quad \mbox{($D_{ij}<0$)}\end{aligned}\right.
\label{averaging1bq}
\end{align}

\begin{figure}[ht]
\includegraphics[width=0.9\linewidth]{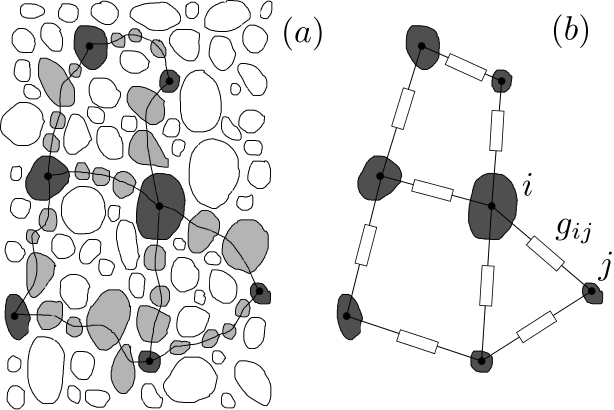}
\caption{a) Sample of granular material. Resonant grains are shown as dark shapes. Electron cotunneling paths between resonant grains go through grey intermediate grains. b) Equivalent Miller-Abrahams network of conductances.}
\label{miller2}
\end{figure}

Thus, we have come to  a system of grains, each of them being characterised by the position ${\bf r}_i$ of it's centre, by the energy $\varepsilon_i$, and by the barrier $W_i$. Obviously, at low $T$ hopping electrons prefer to ``make stops'' only at resonant grains -- those with small $\varepsilon$ and $W$, while the nonresonant grains with typically large $\varepsilon$ and $W$ may serve only as intermediate places, where electrons occur only virtually, staying there for very short times, governed by the quantum uncertainty, see Fig. \ref{miller2}. The random variables $\varepsilon_i$ and $W_i$ have the distribution function $\tilde{\nu}(\varepsilon,W)$ and may be correlated or uncorrelated, depending on the underlying physics. In this paper we will consider only the noncorrelated case, in which for small enough $\varepsilon$ 
\begin{align}
 \tilde{\nu}(\varepsilon,W)= \nu_0P(W)
\label{poissonio}
\end{align}
where $P(W)$ is the barrier distribution function and $\nu_0$ is the density of states (DOGS) near $\varepsilon=0$
\begin{align}
\nu_0=n_g\tilde{P}(\gamma=\pm 1/2)/E_{c}\label{result6}
\end{align}
where $\tilde{P}(\gamma=\pm 1/2)$ is the probability density to have $\gamma=1/2$ (the same as $\gamma=-1/2$), and $n_g$ is the concentration of grains.

\subsection{Statistics of barriers\label{Statistics of barriers}}

Within the paradigm of short range interaction it is natural to assume that the interaction between grains and impurities is also a short range one. It means that only the impurities, situated immediately at the surface of particular grain $i$ contribute to both $\varepsilon_i$ (through the offset charge $\gamma_i$) and $W_i$ (through finite elasticity $\alpha_i$). 

In general, the number of such impurities $p_i$ fluctuates from grain to grain. Different impurities give to $\varepsilon_i$ contributions of different signs, depending on their charges and positions. Therefore $\varepsilon_i$ would fluctuate from grain to grain already in the hypothetical case, where all $p_i$ are the same. 

Because of the random signs of different contributions to $\varepsilon_i$ the density of states $\nu_0(\varepsilon)$ can easily be nonzero at $\varepsilon=0$. 
However, it is not the case for the barriers $W_i$. Each barrier is additive with respect to different impurities
\begin{align}
W_i=\sum_{k_i=1}^{p_i}W_{k_i}
\label{poisson0}
\end{align}
Since the individual contributions are positively defined, the random charges of different impurities do not matter, and the distribution function $P(W)$ should vanish at $W=0$. At the same time, for any positive barrier height $P(W)$ should be nonzero: there is no physical reason to expect a hard gap in $P(W)$. 

\subsubsection{Gaussian distribution of barriers}

The contributions of individual impurities $W_{k_i}$ are assumed to be noncorrelated random variables with identical relatively narrow distributions. Thus, for the most interesting case, in which the average number of impurities coupled to each grain is large $\overline{p}_i\gg 1$, the central part of the distribution $P(W)$ should be a Gaussian one:
\begin{align}
P(W)=\frac{1}{\sqrt{2\pi}\delta W}\exp\left\{-\frac{(W-\overline{W})^2}{2\delta W^2}\right\}
\label{gausian}
\end{align}
where $\overline{W}$ is the average barrier, $\delta W$ is its standard deviation, $\delta W^2\equiv \overline{(W-\overline{W})^2}$. We will stick to the case $\overline{p}\gg 1$, when $\overline{W}\propto \overline{p}$ and $\delta W\propto \overline{p}^{1/2}$, so that the distribution is  narrow: $\delta W\ll \overline{W}$.

The condition $\overline{p}_i\gg 1$ also allows to neglect the correlations between $\varepsilon_i$ and $W_i$. Note that for $p_i\sim 1$ the correlations are strong: for instance, the absence of impurities, associated with given grain (i.e., $p_i=0$) leads to $W_i=0$ and, simultaneously, $\gamma_i=0$.

Thus, in the present paper we are going to analyze the case where the joint density of states \eqref{poissonio} has the polaronic factor $P(W)$ defined in \eqref{gausian}. We have physically justified such choice, although other distribution functions can also be considered. 

\subsubsection{Rectangular distribution of barriers}

For example, in the paper \cite{foygel} a factorized rectangular distribution 
\begin{align}
 P(W)=\frac{\theta(W-W_{\min})\theta(W_{\max}-W)}{W_{\max}-W_{\min}}
\label{rect}
\end{align}
was studied by the method similar to the original percolational approach to the Mott VRH, proposed in \cite{halperin}. The most spectacular result was obtained for the case $W_{\min}\ll\overline{\varepsilon}(T)\ll W_{\max}$,  when for relevant $\varepsilon\sim\overline{\varepsilon}(T)$ within the Mott strip of width $\overline{\varepsilon}(T)$ the``generalized density of states'' $\tilde{\nu}(\varepsilon,W)=\tilde{\nu}_0$ may be treated as $\varepsilon$- and $W$-independent. Under this condition the temperature dependence of conductivity is given by  
\begin{align}
 \sigma\propto\exp\{-(T_M/T)^{2/(d+2)}\},\\ T_M\sim(\tilde{\nu}_0a^d)^{-1/2},\quad \overline{\varepsilon}(T)\sim T^{d/(d+2)}T_M^{2/(d+2)}
 \label{quasi-mott}
\end{align}
For low temperatures, such that $W_{\min}> \overline{\varepsilon}(T)$ the low-energy gap in the distribution \eqref{rect} becomes essential and the conductivity acquires a hard gap as well:
\begin{align}
 \sigma\propto\exp\{-W_{\min}/T\},\quad T\ll W_{\min}\left(\frac{W_{\min}}{T_M}\right)^{\frac{1}{d+1}}
 \label{quasi-mott1}
\end{align}
As we have already noted, in reality the gap in the distribution $P(W)$ is not hard, it is most likely to have an exponential tail at zero.  In the rest of this paper we will adopt the gaussian distribution of barriers \eqref{gausian}.

\subsection{Percolation problem}

The conductivity of Miller-Abrahams network with exponentialy wide distribution of conductances should be found by means of percolation theory \cite{halperin,ES_book}. 

We have come to a kind of coloured percolation problem on random sites, homogeneously distributed in $({\bf r},\epsilon)$ space with the density (DOS) $\nu_0$. Each site $i$, besides its position $({\bf r}_i,\epsilon_i)$, is characterised by a positive variable $W_i$, distributed according to \eqref{gausian}. The variables $\epsilon_i$ and $W_i$ are interpreted as two components of a composite colour of a site $i$, and $n(\epsilon,W)\equiv\nu(\epsilon) P(W)$ is the density of sites with given colour. 

By definition, a pair of sites $\langle ij\rangle$ is ``$\xi$-connected'' if
\begin{align}
\xi_{ij}\equiv \frac{2|{\bf r}_i-{\bf r}_j|}{a}+\frac{E_{ij}}{T}<\xi
\label{poiss2}
\end{align}
where
\begin{align}
E_{ij}\equiv E(\varepsilon_iW_i|\varepsilon_jW_j)=\varepsilon_{ij}+\Lambda_{ij} 
\label{crit-perc} 
\\
\Lambda_{ij}\equiv\frac{D_{ij}^2}{16W_{ij}}\theta(-D_{ij})
\\
D_{ij}\equiv |\varepsilon_i-\varepsilon_j|-4W_{ij}, \quad W_{ij}\equiv W_{i}+W_{j}
\label{poiss3}
\end{align}

As it is usual for the problems of VRH type, the dc charge transfer processes at low temperatures are confined to certain critical subnetwork of resonant grains with small energies $\varepsilon$ within certain narrow strip of width $\overline{\varepsilon}(T)$. This width is $T$-dependent and has to be defined self-consistently. It is more convenient to proceed with calculations in different temperature ranges separately.

\section{Conductivity in different temperature ranges\label{Conductivity  in different temperature ranges}}

There are three principal energy scales in our problem: (i) The average barrier $\overline{W}$, (ii) The barrier dispersion $\delta W\ll \overline{W}$, and (iii) the width of the Mott strip $\overline{\varepsilon}(T)$. Correspondingly, there are three ranges of temperature with different dominating physics.

\subsection{Electron Hopping -- standard Mott VRH}

In this case $\overline{W}\ll\overline{\varepsilon}(T)$, so that the polaron effect is negligible, and the standard Mott law is valid
\begin{align}
\sigma\propto \exp\left\{-\left(\frac{T_M}{T}\right)^{\frac{1}{d+1}}\right\},\qquad T_M=\frac{\beta_M}{\nu_0a^d}
\label{vott1n}
\end{align}
The width of effective energy strip is
\begin{align}
\overline{\varepsilon}(T)\sim T\left(T_M/T\right)^{\frac{1}{d+1}}
\end{align}
The condition $\overline{W}\ll\overline{\varepsilon}(T)$ is equivalent to 
\begin{align}
T\gg T_{c1},
\qquad T_{c1}\sim \overline{W}\left(\frac{\overline{W}}{T_M}\right)^{\frac{1}{d}}
\label{vott2nn}
\end{align}
being the temperature of crossover  between the electron and the polaron regimes.

\subsection{Polaron Hopping, grains with typical barriers}

Here 
\begin{align}
\delta W\ll\overline{\varepsilon}(T)\ll \overline{W}
\label{coon2}
\end{align}
so that the polaron effect is dominant, but the fluctuations of the barriers are still negligible. In this range the results of the model with identical barriers for all grains are applicable:
\begin{align}
\sigma\propto \exp\left\{-\frac{2\overline{W}}{T}-\left(\frac{T_M'}{T}\right)^{\frac{1}{d+1}}\right\},\nonumber\\ T'_M=\frac{\beta'_M}{\nu_Fa^d}, \qquad 
\overline{\varepsilon}(T)\sim T\left(\frac{T'_M}{T}\right)^{\frac{1}{d+1}}
\label{vott1}
\end{align}
Note that the second term in the exponent is small compared to the first one.
The condition $\delta W\ll\overline{\varepsilon}(T)\ll \overline{W}$ is equivalent to $T_{c2}\ll T\ll T_{c1}$, where
\begin{align}
T_{c2}\sim \delta W\left(\frac{\delta W}{T'_M}\right)^{\frac{1}{d}}\sim T_{c1} \left(\frac{\delta W}{\overline{W}}\right)^{\frac{d+1}{d}}\ll T_{c1}
\label{vott2}
\end{align}
In principle, one can also write a general formula that describes the behaviour of the conductivity in the entire range $T\gg T_{c2}$: 
\begin{align}
\sigma\propto \exp\left\{-\frac{W_{\rm eff}(T)}{T}\right\},\quad W_{\rm eff}(T)=\overline{W}\left[2+F_1(T/T_{c1})\right]
\label{vott1vf}
\end{align}
$F_1(x)$ being  a universal function, with a shape, depending only on the space dimensionality $d$. Its asymptotics are
\begin{align}
 F_1(x)\approx x^{\frac{d}{d+1}}\left\{\begin{aligned}
 c_1,& \qquad(x\ll 1)\\
 c_2,& \qquad(x\gg 1)
 \end{aligned}\right.
 \label{asym1}
\end{align}
where $c_1,c_2$ are some universal constants. An interpolation formula for the function $F_1(x)$ was proposed in \cite{foygel}.

\subsection{Polaron Hopping, grains with the barriers in the Gaussian tail of distribution}

Here 
\begin{align}
\overline{\varepsilon}(T)\ll \delta W\ll\overline{W}
\label{coon33}
\end{align}
 and the grains with anomalously low barriers $W_i$ from the tail of the distribution \eqref{gausian} dominate the critical network. 
 \begin{align}
\sigma\propto\exp\left\{-\frac{2\overline{W}}{T}+\frac{\delta W}{T}\left[8d\ln\left(\frac{T_{c2}}{T}\right)\right]^{1/2}\right\},\nonumber\\
\overline{\varepsilon}(T)\sim\delta W\left[8d\ln\left(\frac{T_{c2}}{T}\right)\right]^{-1/2}
\label{votte2}
\end{align}
Namely, the critical subnetwork consists of grains with $W_i$ from a narrow strip
\begin{align}
\left|W_i- \frac12W_{\rm eff}(T)\right|\lesssim \overline{\varepsilon}(T),\nonumber\\
W_{\rm eff}(T)\equiv2\overline{W}-\delta W\left[8d\ln\left(\frac{T_{c2}}{T}\right)\right]^{1/2}
\label{coon3}
\end{align}
The derivation of results \eqref{votte2} and \eqref{coon3} is given in the Appendix \ref{deriva}. It is based on the possibility to reduce the initial multiparametrical percolation problem to certain universal one, governed by a single parameter.

Note, that the result \eqref{votte2} is only valid, if the second term in the exponent is small compared to the first one. In the terms of temperature it means
\begin{align}
T_{{\min}}\ll T\ll T_{c2},\qquad T_{\min}\sim T_{c2}\exp\left\{-\frac{1}{8d}\left(\frac{\overline{W}}{\delta W}\right)^2\right\}
\label{coon3s}
\end{align}
In principle, one can also combine the results in the entire range $T_{\min}\ll T\ll T_{c1}$ in one formula:
\begin{align}
\sigma\propto \exp\left\{-\frac{W_{\rm eff}(T)}{T}\right\},\quad W_{\rm eff}(T)=2\overline{W}+\delta W F_2(T/T_{c2})
\label{vott1vf1}
\end{align}
where $F_2(x)$ is, again, a universal function, depending only on the space dimensionality $d$. Its asymptotics are
\begin{align}
 F_2(x)\approx\left\{\begin{aligned}
 -\sqrt{8d\ln(1/x)},& \qquad(x\ll 1)\\
 c_3x^{\frac{d}{d+1}},& \qquad(x\gg 1)
 \end{aligned}\right.
 \label{asym2}
\end{align}
where $c_3$ is a constant. Note that the high-temperature asymptotics of \eqref{vott1vf1} coincides with the low-temperature asymptotics of \eqref{vott1vf}.

The inequality $T\gg T_{\min}$ was imposed to secure the condition of relatively small relevant fluctuations with $W-\overline{W}\ll \overline{W}$. This condition is needed to justify their universal gaussian distribution. At $T\sim T_{\min}$ the relevant fluctuations are large: $W-\overline{W}\sim\overline{W}$ and their distribution is strongly model-dependent.

The range of exponentially low temperatures $T\lesssim T_{\min}$ may only be of academic interest, because at such low $T$ the the barrier would rather be penetrated due to quantum tunneling mechanism, than due to the activational one, which we are discussing here.

\subsection{Effective radius of the wavefunctions}

Now we have to specify the ``effective radius of the wave function'' $a$, entering the criterion \eqref{poiss2}.
In the Electron Hopping regime, when $|\varepsilon| \gg W$, the actual ``radius'' $a=a_{ij}$ depends on the characteristics of particular grains $i$ and $j$. This logarithmic dependence can be easily  taken into account. This was done in \cite{iosel2005} with the help of the perturbational method in the percolation theory. As a result, the effective radius $a$ appeared to be inversely proportional to the large logarithm ${\cal L}(T)$:
\begin{align}
a=a(T)=\frac{2a_0}{{\cal L}(T)}
\end{align}
where we define $a_0$ as the mean distance between the grains: $r_{ij}=a_0{\cal N}_{ij}$. As a consequence, the parameter $T_M$ in \eqref{vott1n} acquires a logarithmic temperature dependence
\begin{align}
T_M=T_M(T)=\frac{\beta'}{\nu_Fa(T)^d}=\frac{\beta}{\nu_F}\left(\frac{{\cal L}(T)}{2a_0}\right)^d
\label{a0}
\end{align}

Fortunately, for $|\varepsilon| \ll W$, in all the Polaron Hopping regimes,  including both weak fluctuations and strong fluctuations cases,  $a$ doesn't depend on $i$ and $j$, but only on temperature. This can be seen from the expression for transition rate, which contains the factor
\begin{align}
e^{-2r_{ij}/a}\equiv\left\{\frac{A_1g\delta}{E_c}\left[1+\left(\frac{\pi T}{T_*\sin\frac{\pi|D|}{8W}}\right)^2\right]\right\}^{{\cal N}_{ij}}\approx
\nonumber\\
\approx\left\{\frac{A_1g\delta}{E_c}\left[1+\left(\frac{\pi T}{T_*}\right)^2\right]\right\}^{{\cal N}_{ij}}
\end{align}    
Thus, no perturbational method is needed in this case and 
\begin{align}
a=a'(T)=\frac{2a_0}{\ln\left(\frac{E_c}{A_1g\delta}\right)-\ln\left[1+\left(\frac{\pi T}{T_*}\right)^2\right]}
\label{a}
\end{align}
so that the parameter $T_M'$ in \eqref{vott1} also becomes logarithmically $T$-dependent:
\begin{align}
T_M'=T_M'(T)=T_M'(0)\left(1-\frac{\ln\left[1+\left(\frac{\pi T}{T_*}\right)^2\right]}{\ln\left(\frac{E_c}{A_1g\delta}\right)}\right)^d, \nonumber
\end{align}
\begin{align}
T_M'(0)=\frac{\beta'}{\nu_Fa'(0)^d}=\frac{\beta'}{\nu_F}\left(\frac{\ln\left(\frac{E_c}{A_1g\delta}\right)}{2a_0}\right)^d
\label{a1}
\end{align}

%It is hardly possible to find $a$ in the case when $|\varepsilon|\sim W\sim \xi T$. Inspecting the expressions for $a(T)$ in EH regime and the formula \eqref{a} for PH regime, we can guess that in general we have
%\begin{align}
%a=\frac{2a_0}{\ln\left\{\frac{E_c}{A_1g\delta}\left[1+\left(\frac{\alpha T}{T_*}\right)^2\right]^{-1}\right\}}
%\label{a1}
%\end{align}
%where $\alpha\equiv\alpha(T)$ is the dimensionless function of temperature. The crossover from elastic to inelastic cotunneling happens at temperature $T_*/\alpha$. In the range of temperatures of the PH regime we have $\alpha=\pi$, while for the EH regime we know that elastic-inelastic crossover happens at $T_*/\cal L$, and thus $\alpha\sim {\cal L}$. In the intermediate region, where $|\varepsilon|\sim W \sim \xi T$, we only know that $\alpha$ smoothly changes from $\pi$ to ${\cal L}$.

%\subsection{Differential activation energy}

%Experimentalists often introduce the  differential activation energy according to
%\begin{align}
%E_{\rm act}(T)\equiv\frac{\partial \ln (1/\sigma(T))}{\partial (1/T)}
%\label{effact}
%\end{align}
%For our problem we get
%\begin{align}
%E_{\rm act}\approx\left\{\begin{aligned}
%\frac{T}{d+1}\left(\frac{T_M(T)}{T}\right)^{\frac{1}{d+1}}&,\quad (T_{c1}\ll T),\\
%\overline{W}+\frac{T}{d+1}\left(\frac{T'_M(T)}{T}\right)^{\frac{1}{d+1}}&,\quad (T_{c2}\ll T\ll T_{c1}),\\
%\overline{W}-\delta W\left[2d\ln\left(\frac{T_{c2}}{T}\right)\right]^{1/2}&,\quad (T_{\min}\ll T\ll T_{c2}).
%\end{aligned}
%\right.
%\label{effact1}
%\end{align}

\section{Summary \label{Summary}}

We have considered  a granular metal with charged impurities (stray charges) that are not rigidly fixed at certain positions, but can be slightly displaced from their equilibrium positions due to electrostatic forces and thermal fluctuations. The flexibility of the system of impurities leads to a ``polaronic effect'', that can be either weak (for $T\gg T_{c1}$) or strong (for $T\ll T_{c1}$), where the crossover temperature 
\begin{align}
T_{c1}\sim \overline{W}\left(\overline{W}/T_M\right)^{\frac{1}{d}}
\label{vott2nn1}
\end{align}
and $T_{M}\approx T_{M}(T_{c1})$ is given by \eqref{a0wwqa} below. 

In this paper we have studied a model with short range interaction (e.g., screened Coulomb) and taken into account only the thermoactivational mechanism of the penetration of the polaronic barrier by the configurational degrees of freedom, not the tunneling one. The latter may become relevant at very low temperatures (the larger the effective mass $M$ of the impurities, the lower the corresponding crossover temperature). 

Experimentalists often introduce the  differential activation energy according to
\begin{align}
E_{\rm act}(T)\equiv\frac{\partial \ln (1/\sigma(T))}{\partial (1/T)}
\label{effact}
\end{align}
Below we discuss the temperature dependence of $E_{\rm act}(T)$ in different temperature ranges.

\subsection{Weak polaronic effect}

In the range $T\gg T_{c1}$ the  hopping can be with minor modifications described by the electronic multiple cotunneling \cite{iosel2005}
\begin{align}
E_{\rm act}(T)\approx
\frac{T}{d+1}\left(\frac{T_M(T)}{T}\right)^{\frac{1}{d+1}}
\label{effact1a}
\end{align}

\subsection{Strong polaronic effect}

The range $T\ll T_{c1}$, where the cotunneling is dominated by polaronic effect, is split into two subranges with the crossover temperature
\begin{align}
T_{c2}\sim  T_{c1} \left(\delta W/\overline{W}\right)^{\frac{d+1}{d}}\ll T_{c1}
\label{vott2df}
\end{align}

\begin{itemize}
\item For $T_{c2}\ll T\ll T_{c1}$ the spatial fluctuations of polaronic barriers $W_i$ can be neglected and the choice of resonant grains that constitute the effectively conducting network is  dictated exclusively by the values of $\varepsilon_i$: they should lie in the Mott-like strip $|\varepsilon_i|\lesssim  \overline{\varepsilon}(T)$, while the particular values of $W_i$ are irrelevant. Here the approach, similar to the one, proposed in \cite{foygel}, is applicable, and
\begin{align}
E_{\rm act}(T)\approx
\overline{W}+\frac{T}{d+1}\left(\frac{T'_M(T)}{T}\right)^{\frac{1}{d+1}}
\label{effact1b}
\end{align}
where the second term is relatively small, compared to the first one.

\item  For $ T\ll T_{c2}$ both $\varepsilon_i$ and  $W_i$ are important, the effective network is formed by the grains for which both $\varepsilon_i$ and  $W_i$ are anomalously small: $|\varepsilon_i|\lesssim  \overline{\varepsilon}(T)$ and simultaneously $|W_i-\frac12W_{\rm eff}(T)|\lesssim  \overline{\varepsilon}(T)$. Here
\begin{align}
\overline{\varepsilon}(T)\sim\delta W\left[8d\ln\left(\frac{T_{c2}}{T}\right)\right]^{-1/2}
\label{effact1cfd}
\end{align}
In this temperature range
\begin{align}
E_{\rm act}(T)\approx W_{\rm eff}\equiv
2\overline{W}-\delta W\left[8d\ln\left(\frac{T_{c2}}{T}\right)\right]^{1/2}
\label{effact1c}
\end{align}
\end{itemize}

Thus, due to the tail in the distribution of the barrier fluctuations, the effective activation energy continues to decrease even at lowest temperatures, though this decrease becomes very slow.

\subsection{Elastic vs inelastic cotunneling}

The above mentioned crossovers discriminate different modes of hopping with respect to the strength of polaronic effect (crossover at $T\sim T_{c1}$) and to the importance of the fluctuations of the latter (crossover at $T\sim T_{c2}$). There is, however, one additional crossover at $T\sim T_{c0}$ that discriminates different modes with respect to the character of cotunneling: elasic or inelastic. For $T\gg T_{c0}$ the ``constants'' $T_{M}(T)$ and $T'_{M}(T)$ logarithmically increase with the lowering of temperature, while at $T\lesssim T_{c0}$ they saturate and become $T$-independent:
\begin{align}
T_M(T)=\frac{\beta{\cal L}^d(T)}{\nu_F(2a_0)^d},\quad T'_M=\frac{\beta'{\cal L}^d(0)}{\nu_F(2a_0)^d}
\label{a0wwqa}
\end{align}
\begin{align}
{\cal L}(T)\approx 
\ln\left(\frac{E_c}{g\delta}\frac{T_{c0}^2}{T_{c0}^2+T^2}\right)
\label{essence1}
\end{align}
 The explicit form of $T_{c0}$ depends on the relation between $T_{c0}$ and $T_{c1}$:
 \begin{align}
T_{c0}\sim(E_c\delta)^{1/2}\left\{\begin{aligned}1/{\cal L}(0),&\quad T_{c0}\gg T_{c1}\\
1/\pi,&\quad T_{c0}\ll T_{c1}
\end{aligned}
\right.
\label{essence11}
\end{align}

For $T_{c0}\gg T_{c1}$ the crossover between elastic and inelastic cotunneling takes place within the weak polaron effect domain, 
while for $T_{c0}\ll T_{c1}$ it happens within the strong polaron effect domain.

\section{Conclusion\label{Conclusion}}

In this paper we have introduced a concept of polaronic effect in granular systems, related to the flexibility of the random charges, trapped in the insulating matrix. We have explained, how this effect is manifested in the conductivity of the system, the latter being controlled by multiple cotunneling of electrons through long ``strings'' of adjacent grains. The basic line of our reasoning was similar to that of every VRH-like calculation, and could be split into two basic steps: 

(i) We  calculate the carrier transition rate between distant resonant sites, taking into account all the necessary physics (the cotunneling and coupling to flexible impurities). The transition rate is given by the expression \eqref{large3} and its simplifications \eqref{fin2v} and \eqref{large3dd}. 

(ii) Afterwards, we find the conductivity of Miller-Abrahams network of conductances with the help of percolation theory.  We distinguish two important temperature ranges, namely $T>T_{c1}$ (Electron Hopping) and $T<T_{c1}$ (Polaron Hopping), which differ in both physical characteristics of transport and $T$-dependence of conductivity.   

Besides the crossover between Electron and Polaron hopping we have also studied another crossover -- between elastic and inelastic cotunneling regimes. It takes place at   the temperature $T=T_{c0}$, given by \eqref{essence11}. In the presence of strong polaronic effect this temperature turns out to be the same as for single quantum dot, in contrast with the Electron Hopping regime, where it is much lower.

A few important questions remain open and are subject to future research. 

(i) In the present work we have treated the configurational degrees of freedom (i.e., coordinates of the charged impurities) as classic ones. This can be justified only if the temperature $T$ is higher than characteristic frequencies $\omega_i$ of  the impurities vibrations; at lower  $T$ the polaronic barriers would be penetrated by means of quantum tunneling. In principle, it should lead to the reentrance of Mott law at $T<\overline{\omega}$. However, since $\omega_i$ may be different for different impurities, the crossover to tunneling may occur not simultaneously at all grains, and that may give rise to some interesting new physics.

(ii) We have studied here only the case of short range interactions. Some qualitative ideas about the effects of the long range Coulomb interactions for the Electron Hopping were presented in \cite{iosel2005}.
 
(iii) The lack of experimental data on polaron effect in granular systems does not allow to reliably choose the  distribution function of  the polaronic barriers $W$. It is not clear whether it should be  wide or narrow, have a power-law tail at $W\to 0$ or exponential one. 

(iv) It is also unclear if the present model, in which the occupation numbers $n$ are coupled to oscillators, is adequate for real world applications. Another possible option would be the coupling to the tunneling two-level systems (TTLS, see \cite{TTLS1,TTLS2}).  Physically the two-level system may be represented by impurity atoms that can tunnel between two adjacent potential wells. Such systems play important role in physics of glasses, they also contribute to dephasing in qubits \cite{TLS}.

We believe that the formalism developed in this paper will allow us to answer at least the first question from the list  above. As for the other questions, some novel approaches may be required.

We are indebted to M.V.Feigel'man and  L.B.Ioffe for helpful discussions. This work was supported by 5top100 grant of Russian Ministry of Education and Science.

%%%%%%%%%%%%%%%%%%%%%%%%%%%%%%%%%%%%%%%%%%%%%%%%%%%%%%%%%%%%%%%%%%%%%%%%%%%%%%%%%%%%%%%%%%%%%%%%%%%%%%%%%%%%%%%%%%%%%%

%%%%%%%%%%%%%%%%%%%%%%%%%%%%%%%%%%%%%%%%%%%%%%%%%%%%%%%%%%%%%%%%%%%%%%%%%%%%%%%%%%%%%%%%%%%%%%%%%%%%%%%%%%%%%%%%%%%%%%

\appendix
\section{Percolation with gaussian distribution of barriers. Derivation of the formula \eqref{votte2} \label{deriva}}

We are interested in the low-$T$ behaviour of the system, when the critical subnetwork, responsible for the conductivity, is comprised  of the rare grains with anomalously small $\epsilon_i$ and anomalously low barriers $W_i$ from the tail of the distribution \eqref{gausian}. As we will see later, these parameters are confined in narrow strips
\begin{align}
\epsilon_i\lesssim\overline{\varepsilon}_{\epsilon},\qquad W_i-\xi T\lesssim\overline{\varepsilon}_{W},
\qquad \mbox{with}\;\; \overline{\varepsilon}_{\epsilon},\overline{\varepsilon}_{W}\ll\xi T
\label{strip1}
\end{align}
 Now we will adopt the inequalities \eqref{strip1} as a conjecture, the  real values of $\overline{\varepsilon}_{\epsilon},\overline{\varepsilon}_{W}$ will be determined later, and these values will justify \eqref{strip1}.
 
 Based on the assumption \eqref{strip1}, the expression \eqref{poiss3} for $\Lambda_{ij}$ can be expanded and we get
\begin{align}
 E(\epsilon_iW_i|\epsilon_jW_j)=\frac{|\epsilon_i|+|\epsilon_j|}{2}+W_i+W_j
\label{poiss3x}
\end{align}
The result \eqref{poiss3x} means that indeed each site may be characterised by one ``composite colour'' $\tilde{W}_i\equiv W_i+|\epsilon_i|/2>0$ not by two independent colours $\epsilon_i$ and $W_i$. The density of this composite colour is
\begin{align}
 \tilde{n}(\tilde{W})=2\int_0^{2\tilde{W}}n(\epsilon,\tilde{W}-\epsilon/2)d\epsilon 
\label{strip2}
\end{align}
(the factor 2 arises due to symmetry $\varepsilon\to-\varepsilon$), and 
\begin{align}
\xi_{ij}\equiv \frac{2|{\bf r}_i-{\bf r}_j|}{a}+\frac{E_{ij}}{T},\quad E_{ij}\equiv  E(\tilde{W}_i|\tilde{W}_j)=\tilde{W}_i+\tilde{W}_j
\label{poiss3y}
\end{align}
Now we introduce dimensionless variables
\begin{align}
{\bf x}_i\equiv\frac{2{\bf r}}{\xi a},\qquad \delta_i\equiv\frac12-\frac{\tilde{W}_i}{\xi T}
\label{poiss7s}
\end{align}
and arrive at the dimensionless percolation problem 
\begin{align}
\tilde{\xi}_{ij}\equiv |{\bf x}_i-{\bf x}_j|-\delta_i-\delta_j<0
\label{poiss2v}
\end{align}
with the density in the $({\bf x},\delta)$-space
\begin{align}
\tilde{n}(\delta)=2\xi T\left(\frac{\xi a}{2}\right)^d\int_0^{\xi T\left(\frac12-\delta\right)}n\left[\epsilon,\xi T\left(\frac12-\delta\right)-\frac{\epsilon}{2}\right]d\epsilon\approx\nonumber\\ \approx  \frac{4\beta'T}{T'_M}(\xi /2)^{d+1}\Phi\left[\frac{\overline{W}-\xi T\left(\frac12-\delta\right)}{\delta W}\right]
\end{align}
where 
\begin{align}
\Phi(x)\equiv\frac{1}{\sqrt{2\pi}}\int_{x}^{\infty}\exp\left\{-\frac{\zeta^2}{2}\right\}d\zeta
\label{strip2axx}
\end{align}
In particular, we will need the asymptotics
\begin{align}
 \Phi(x)\approx\
 \frac{1}{x\sqrt{2\pi}}\exp\{-x^2/2\}, \qquad(x\gg 1)
 \label{strip2aa}
\end{align}
It is convenient to write
\begin{align}
\sigma\propto\exp\left\{-\frac{W_{\rm eff}(T)}{T}\right\},\qquad W_{\rm eff}(T)\equiv \xi T
\label{sigma-gaussq}
\end{align}
and note that $w\equiv 2\overline{W}-W_{\rm eff}(T)\ll \overline{W}$ and $\xi T\delta\ll \overline{W}$, so that
\begin{align}
 \tilde{n}(\delta) \approx
 \frac{4\beta'T}{T'_M}\left(\frac{\xi }{2}\right)^{d+1}\Phi\left(\frac{w/2+2\overline{W}\delta}{\delta W}\right),\nonumber\\
 \Phi\left(\frac{w/2+2\overline{W}\delta}{\delta W}\right)\approx
 \frac{1}{\sqrt{2\pi}}\frac{\delta W}{w/2+2\overline{W}\delta}\times\nonumber\\ \times
 \exp\left\{-\frac{(w/2)^2}{2\delta W^2}-\frac{w}{\delta W^2}\overline{W}\delta-\frac{(2\overline{W}\delta)^2}{2\delta W^2}\right\}
 \label{strip2au}
\end{align}
As we will see soon, for typical $\delta$ and $w$ a hierarchy  $\overline{W}\delta\ll \delta W\ll w$ holds. Therefore only the first two terms in the exponent should be kept, while last term is much less than unity and can be neglected. As a result, we can rewrite \eqref{strip2au} in a form 
\begin{align}
 \tilde{n}(\delta) \approx
 \frac{4\beta'T}{T'_M}\left(\frac{\xi }{2}\right)^{d+1}
 \frac{1}{\sqrt{2\pi}}\frac{2\delta W}{w}\times\nonumber\\ \times
 \exp\left\{-\frac{w^2}{8\delta W^2}-\frac{w\overline{W}\delta}{\delta W^2}\right\}
 \label{strip2aue}
\end{align}

After the renormalisation of variables
\begin{align}
u_i\equiv \frac{w\overline{W}}{\delta W^2}\delta_i,\qquad {\bf y}_i=\frac{w\overline{W}}{\delta W^2}{\bf x}_i
\label{strip2e1}
\end{align}
we arrive at the universal percolation problem, where the sites of colour $u$ are  randomly distributed with the density
\begin{align}
\tilde{n}(u)\approx
Ae^{-u}\label{denn1}
\end{align}
while the 
percolation criterion reads
\begin{align}
\tilde{\xi}_{ij}\equiv |{\bf y}_i-{\bf y}_j|-u_i-u_j<0
\label{poiss2vw}
\end{align}
Note that this problem is characterised by  single constant 
\begin{align}
A\equiv \left(\frac{w\overline{W}}{\delta W^2}\right)^{-(d+1)}\tilde{n}(\delta=0)=\nonumber\\
\frac{8\beta'T}{T'_M}\frac{1}{\sqrt{2\pi}}\frac{\delta W}{w}\left(\frac{wT}{\delta W^2}\right)^{-(d+1)}\exp\left\{-\frac{w^2}{8\delta W^2}\right\}
\end{align}
and therefore the percolation  should  be established  at 
\begin{align}
A=A_d
\label{poiss2vw1}
\end{align}
where $A_d\sim 1$ is some universal constant, depending only on the space dimensionality $d$.

So, the dependence of the effective barrier $W_{\rm eff}(T)\equiv 2\overline{W}-w$ on the parameters may be found from the equation for $w$:
\begin{align}
1=\frac{8\beta'T}{A_dT'_M}\frac{1}{\sqrt{2\pi}}\frac{\delta W}{w}\left(\frac{wT}{\delta W^2}\right)^{-(d+1)}\exp\left\{-\frac{w^2}{8\delta W^2}\right\}
\label{strip2e2}
\end{align}

In the leading logarithmic approximation the solution of \eqref{strip2e2} reads
\begin{align}
\frac{w}{\delta W}\approx\sqrt{8d\ln(T_{c2}/T)},\quad  T_{c2}\sim\delta W\left(\frac{\delta W}{T'_M}\right)^{1/d}
\label{striffweaq}
\end{align}

Now we can estimate $\overline{\varepsilon}_{\varepsilon}$ and $\overline{\varepsilon}_W$. As it follows from  \eqref{strip2e1}
\begin{align}
\overline{\varepsilon}_{\varepsilon}\sim\overline{\varepsilon}_W\sim \frac{\delta W^2}{w}\sim\frac{\delta W}{\sqrt{8d\ln(T_{c2}/T)}}\ll \delta W
\label{striffweaw1}
\end{align}
which justifies our conjecture \eqref{strip1}.

%%%%%%%%%%%%%%%%%%%%%%%%%%%%%%%%%%%%%%%%%%%%%%%%%%%%%%%%%%%%%%%%%%%%%%%%%%%%%%%%%%%%%%%%%%%%%%%%%%%%%%%%%%%%%%%%%%%%%%

%\appendix

\end{document}